%% file: RANGER-submission.tex
\documentclass{article} 
\usepackage{natbib}
\usepackage{iclr2026_conference,times}
\usepackage{multirow} 
\usepackage{float}
\usepackage{listings}
\usepackage{xcolor}
\usepackage{algorithm} 
\usepackage{array}
\usepackage{booktabs} 
\usepackage{multirow}
\usepackage{siunitx} 
\usepackage{caption}
\usepackage{subcaption}
\usepackage{float}

\usepackage{longtable}
\input{math_commands.tex}

\usepackage{graphicx}
\usepackage{hyperref}
\usepackage{url}
\usepackage{svg}
\usepackage[inline]{enumitem}

\usepackage{amsmath, amssymb}    
\usepackage{algorithm}           
\usepackage[noend]{algpseudocode}

\algrenewcommand\algorithmiccomment[1]{\hfill{\footnotesize\(\triangleright\)\,#1}}

\lstdefinestyle{mystyle}{
    basicstyle=\footnotesize\ttfamily,
    breaklines=true,
    frame=single,
    numbers=left,
    numberstyle=\tiny,
    showstringspaces=false,
    backgroundcolor=\color{gray!10},
}
\definecolor{lightgray}{gray}{0.9}
\definecolor{darkgray}{gray}{0.3}

\title{RANGER: \textbf{R}epository‑level \textbf{A}ge\textbf{n}t for \textbf{G}raph‑\textbf{E}nhanced \textbf{R}etrieval}

\author{
Pratik Shah$^{1,2}$\thanks{Work done during Pratik's internship at Nutanix.}, 
Rajat Ghosh$^{2}$, 
Aryan Singhal$^{2}$, 
Debojyoti Dutta$^{2}$ \\
$^{1}$Georgia Institute of Technology \\
$^{2}$Nutanix \\
\texttt{pratik2002shah@gmail.com, rajat.ghosh@nutanix.com}
}
\iclrfinalcopy 
\begin{document}

\maketitle

\begin{abstract}
General-purpose automated software engineering (ASE) includes tasks such as code completion, retrieval, repair, QA, and summarization. These tasks require a code retrieval system that can handle specific queries about code entities, or \textit{code entity queries} (for example, locating a specific class or retrieving the dependencies of a function), as well as general queries without explicit code entities, or \textit{natural language queries} (for example, describing a task and retrieving the corresponding code). We present \textbf{RANGER}, a repository-level code retrieval agent designed to address both query types, filling a gap in recent works that have focused primarily on code-entity queries. We first present a tool that constructs a comprehensive knowledge graph of the entire repository, capturing hierarchical and cross-file dependencies down to the variable level, and augments graph nodes with textual descriptions and embeddings to bridge the gap between code and natural language. RANGER then operates on this graph through a dual-stage retrieval pipeline. Entity-based queries are answered through fast Cypher lookups, while natural language queries are handled by MCTS-guided graph exploration. We evaluate RANGER across four diverse benchmarks that represent core ASE tasks including code search, question answering, cross-file dependency retrieval, and repository-level code completion. On CodeSearchNet and RepoQA it outperforms retrieval baselines that use embeddings from strong models such as Qwen3-8B.  On RepoBench, it achieves superior cross-file dependency retrieval over baselines, and on CrossCodeEval, pairing RANGER with BM25 delivers the highest exact match rate in code completion compared to other RAG methods.

\end{abstract}

\section{Introduction}

\input{final_sections/introduction.tex}

\section{Related Work}

\input{final_sections/related_work}

\begin{figure}[h]
\centering

\includegraphics[width=1\textwidth]{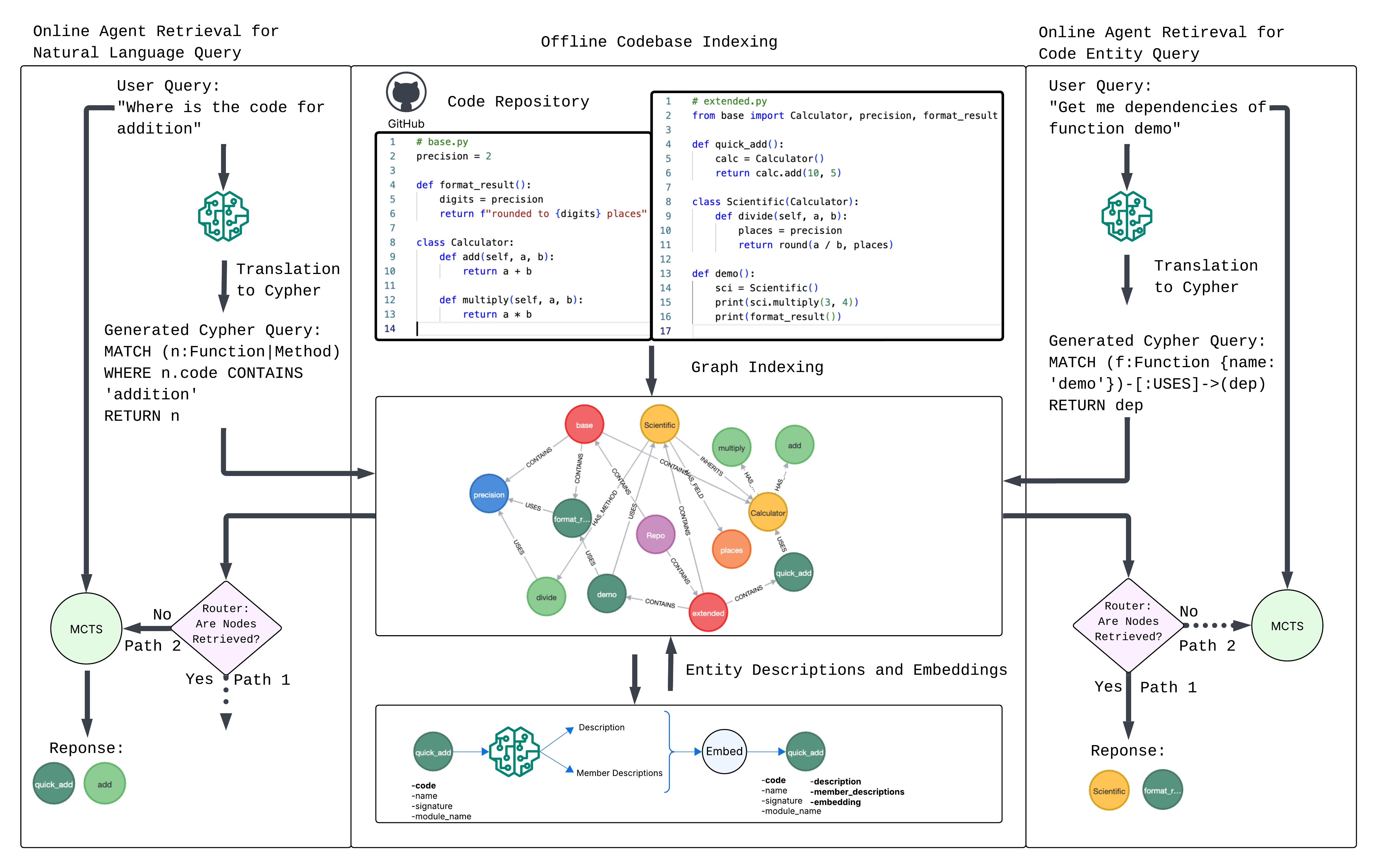}
\caption{RANGER system architecture illustrated through a simple two-file repository example containing \texttt{base.py} and \texttt{extended.py}. The offline stage constructs a comprehensive knowledge graph from code repositories through AST parsing, LLM-assisted semantic description generation, and embedding computation. In the online stage RANGER first translates user queries into Cypher queries using an LLM. For \textit{code-entity queries}, this Cypher query is sufficient and provides fast retrieval (Path 1). If retrieval results from the graph database return \texttt{None}, often in case of \textit{natural language queries}, the system invokes MCTS-based graph exploration (Path 2) to generate the final response.}

\label{fig:overall}

\end{figure}

\section{Methodology}\label{sec:Methodology}

\input{final_sections/methodology.tex}
\section{Experiments}\label{sec:experiments}

We evaluate RANGER on four diverse datasets spanning both \textit{code-entity} and \textit{natural-language} query types and three practical scenarios covering repository-level code retrieval, code completion, and question answering.

\subsection{Natural Language Query Based Retrieval}

\subsubsection{Datasets \& Setup}

\textbf{CodeSearchNet} Challenge (Python split) consists of 99 natural language queries with expert relevance annotations over a large corpus of Python functions \citep{husain2019codesearchnet}. We select 70 repositories with the highest query counts, build knowledge graphs from corresponding commits, and prune nodes not present in the official corpus to align with ground truth annotations.

\textbf{RepoQA} originally evaluates long context code understanding via the Searching Needle Function task where multiple functions are provided to an LLM as context along with a function description and the LLM must return the corresponding function. To facilitate our evaluation we modify the task so that all functions become our corpus and the function description becomes our natural language query \citep{liu2024repoqa}.The function description includes Purpose, Input, Output, and Procedure fields, but to better reflect realistic queries, we use only the Purpose field as the natural language query. We use the Python split with ten repositories and ten descriptions per repository. 

For both datasets we generate text descriptions and embeddings as detailed in Section~\ref{sec:Methodology} and run the MCTS stage for retrieval.

\subsubsection{Baselines and Results}

\begin{figure}[h]
\begin{center}
\includegraphics[width=0.7\textwidth]{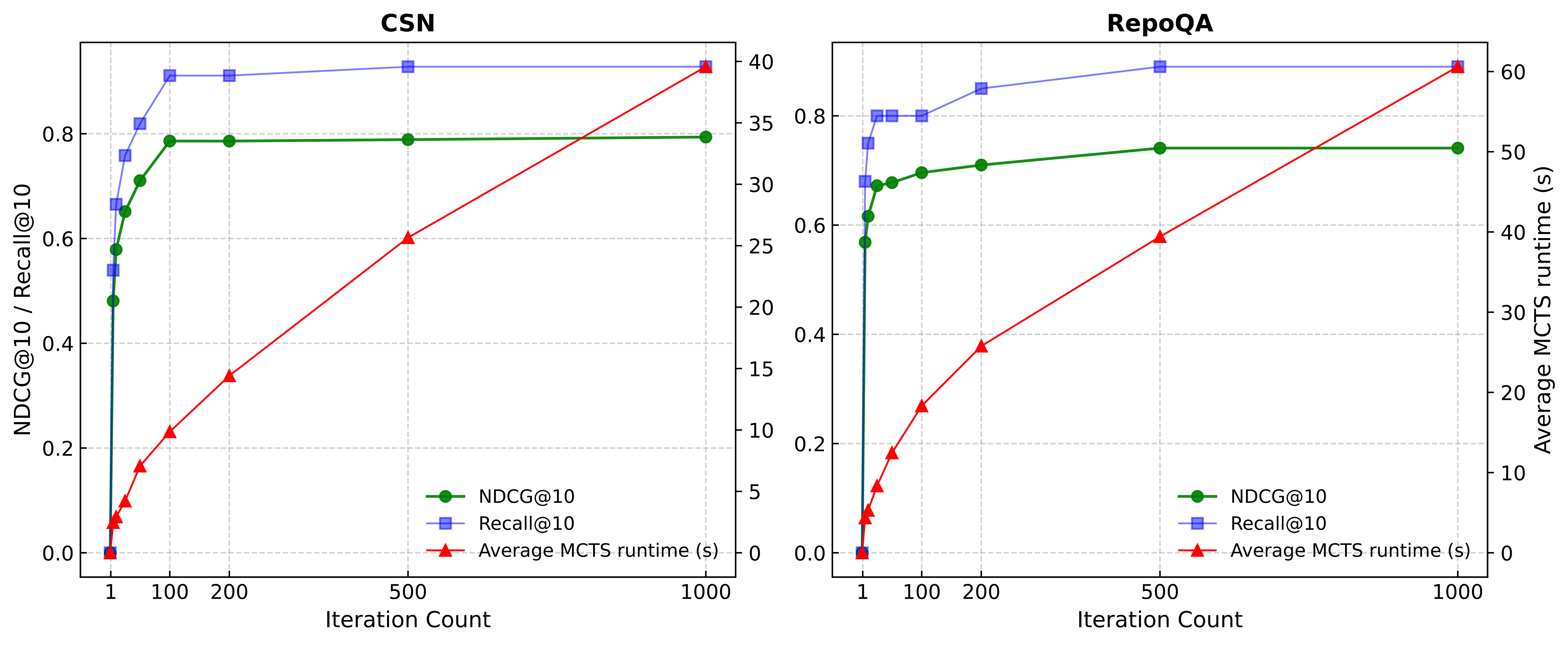}

\end{center}
\vspace{-0.3cm}
\caption{Performance metrics across MCTS iterations for natural language query datasets. Left shows CodeSearchNet NDCG@10, Recall@10, and the average MCTS runtime per iteration across repositories and queries. Right shows RepoQA NDCG@10, Recall@10, and corresponding average runtimes. Both datasets show monotonic improvement followed by convergence, indicating practical iteration ranges for deployment.}

\label{fig:nlq_metrics}
\end{figure}

We compare to two vector search baselines. The first uses raw code embeddings indexed directly from corpus chunks. The second uses embeddings of LLM generated semantic descriptions. This isolates MCTS gains beyond gains from descriptive text.\\ Table~\ref{tab:retrieval_combined} reports NDCG@10 and Recall@10 on CodeSearchNet and RepoQA. RANGER improves both metrics over the baselines and also exceeds retrieval with {Qwen-3-8B} \citep{wang2025qwen3} embeddings which are currently top ranked on the MTEB leaderboard \citep{muennighoff2022mteb}. The improvements stem from the use of cross-encoder scoring, which provides higher accuracy than bi-encoder similarity but is too expensive to apply exhaustively. RANGER addresses this with an MCTS-guided traversal, where the bi-encoder expands promising graph paths and the cross-encoder is applied only to high-value candidates. This selective application preserves the accuracy benefits of cross-encoders while keeping retrieval computationally tractable. \\
Figure~\ref{fig:nlq_metrics} shows that NDCG@10 and Recall@10 improve steadily with additional MCTS iterations before the rate of improvement slows. 
The curves exhibit clear knees that indicate the {optimal iteration range for practical deployment}, balancing retrieval quality with computational cost. 

\begin{table}[t]
\centering
\renewcommand{\arraystretch}{1.3} 
\setlength{\tabcolsep}{10pt} 

\resizebox{\textwidth}{!}{%
\begin{tabular}{l c cc cc}
\toprule
\textbf{Metric} & \textbf{RANGER} & \multicolumn{2}{c}{\textbf{Code Embedding}} & \multicolumn{2}{c}{\textbf{Text Embedding}} \\
 & (MCTS iter) & CodeT5-110M & Qwen-3-8B & Qwen-3-8B & mxbai$^{1}$ (335M) \\
\midrule
\multicolumn{6}{c}{\textbf{CodeSearchNet Dataset}} \\
\midrule
NDCG@10   & \textbf{0.786} (200) & 0.419 & 0.725 & 0.701 & 0.664 \\
Recall@10 & \textbf{0.911} (200) & 0.643 & 0.891 & 0.856 & 0.847 \\
\midrule
\multicolumn{6}{c}{\textbf{RepoQA Dataset}} \\
\midrule
NDCG@10   & \textbf{0.741} (500) & 0.718 & 0.722 & 0.709 & 0.706 \\
Recall@10 & \textbf{0.890} (500) & 0.810 & {0.850} & 0.810 & 0.810 \\
\bottomrule
\end{tabular}}
\caption{\textbf{Performance comparison on CodeSearchNet and RepoQA.} 
RANGER consistently outperforms baseline embedding models across datasets. Iteration counts are shown in parentheses. Best baseline results are bolded.}
\label{tab:retrieval_combined}
\end{table}

\footnotetext[1]{mxbai-embed-large-v1}

\subsection{Code-Entity Query Based Retrieval}

\subsubsection{Dataset \& Setup}
\textbf{RepoBench} \citep{liu2023repobench_python_v11} evaluates repository-level retrieval via RepoBench-R, where the task is selecting the most relevant cross-file snippet to support next-line prediction. We use the Python v1.1 split and restrict to repositories with at least five data points (430 repositories). The prompt provides an incomplete in-file chunk with code entities, which RANGER converts into Cypher queries to retrieve cross-file dependencies before ranking (example in Appendix). Because commit IDs were not released and repositories changed after dataset creation, we use the latest commit as of December 31 2023 and re run baselines for consistency. Since all queries here are code entity queries handled directly by Stage~1 we omit text descriptions which are mainly needed for Path 2 MCTS to reduce compute.

\subsubsection{Baselines and Results}
Following RepoBench-R setup, the baseline treats import statement snippets as candidate contexts which captures file level linkage. Both RANGER and the baseline use the same rerankers and the same top k protocol to isolate retrieval effects.\\
Our graph agent improves Accuracy@5, NDCG@5 and MRR@5 across rerankers which shows better localization of fine grained dependencies than file level imports. Pure semantic retrieval performs poorly which supports the need for cross-file graph traversal over linear index search. See Table~\ref{tab:repobench_result}.
\input{pratik_tables/repobench_result.tex}

\subsection{Code-Entity Query Based Code Completion}

\subsubsection{Dataset \& Setup}
\textbf{CrossCodeEval} \citep{ding2023crosscodeeval} tests cross file code completion across Python, Java, TypeScript and C\# using real repositories where the correct continuation depends on cross file context and not just the current file. We use the Python split with 471 repositories, build knowledge graphs from the dataset specified commits, and retrieve cross file context via RANGER. Same as Repobench, for each repository, a code knowledge graph is constructed from the target commit, which is provided in the datasets, without creating text descriptions. 

\subsubsection{Baselines and Results}
We compare RANGER against BM25 and several repository level retrievers. \textbf{BM25} \citep{robertson2009probabilistic} serves as a strong sparse lexical baseline by selecting top-k contexts via term-frequency scoring. \textbf{CGM MULTI} \citep{tao2025codegraphmodel} constructs a one hop ego subgraph around the active file and applies graph aware attention. \textbf{RepoFuse} \citep{cao2024repofuse} fuses analogy contexts with rationale contexts. \textbf{RLCoder} \citep{wang2024rlcoder} learns a retrieval policy with perplexity based rewards and a learned stopping rule. \textbf{R2C2} \citep{liu2024r2c2coder} assembles repository aware prompts by selecting candidate snippets with context conditioning. Inspired by RepoFuse, which shows that fusing analogy and rationale contexts improves code generation, we also report \textbf{RANGER+BM25} which pairs graph based cross file retrieval with BM25. Since some methods such as RepoFuse and R2C2 use a limit of 4{,}096 tokens on the retrieved context we also present results with a 4{,}096 token limit in the Appendix ~\ref{sec:cceval_appendix}. \\
Table~\ref{tab:retrieval_benchmarking_crosscodeeval} reports Exact Match and Edit Similarity across DeepSeek Coder 7B and CodeLlama 7B. \textbf{RANGER+BM25} achieves the highest Exact Match with DeepSeek Coder 7B and CodeLlama 7B and competitive Exact Match with StarCoder 7B while consistently outperforming BM25. Edit Similarity is mid to strong which reflects the tradeoff between precise dependency localization and lexical similarity. These results underscore the value of explicit graph based retrieval in repository level code completion.
\input{pratik_tables/crosscodeeval_result}

\section{Conclusion}

\input{final_sections/conclusion}

\bibliography{iclr2026_conference}
\bibliographystyle{iclr2026_conference}
\newpage
\appendix
\section{Appendix}
\input{final_sections/appendix}
\end{document}

%% file: math_commands.tex
\usepackage{amsmath,amsfonts,bm}

\def\eqref#1{equation~\ref{#1}}

\def\1{\bm{1}}

\DeclareMathAlphabet{\mathsfit}{\encodingdefault}{\sfdefault}{m}{sl}
\SetMathAlphabet{\mathsfit}{bold}{\encodingdefault}{\sfdefault}{bx}{n}



%% file: final_sections/introduction.tex
Retrieving relevant code snippets, functions, and classes from large repositories is central to modern software engineering, as the quality of retrieved context underpins downstream tasks for AI agents and large language models, including code generation, patch generation, automated program repair, and intelligent code completion. While retrieval over natural language has seen rapid progress \citep{karpukhin2020dense,izacard2022unsupervised}, code retrieval remains substantially more challenging. Unlike natural language, code often contains long-range and multi-hop dependencies \citep{allamanis2018survey}, where the semantics of a program may depend on variables, function calls, or imports that appear far apart in the source. These properties render simple flat indexing insufficient for code retrieval, motivating the use of graph databases \citep{liu2024codexgraphbridginglargelanguage} and multi-hop reasoning to capture cross-file relationships, call graphs, and dependency chains \citep{guo2022unixcoder,ye2022retrieval}. 

An additional challenge in code retrieval arises from query diversity. \textit{Code-entity queries} ask questions about specific code-entities (e.g., “What are the dependencies of \texttt{Calculator} class?”). In contrast, \textit{natural language queries}, describe behaviors or constraints without naming symbols (e.g., “Where do we implement addition?”). Natural language queries \citep{mastropaolo2021empirical,zhang2022codexglue} are particularly difficult due to the semantic gap between natural and symbolic languages \citep{husain2019codesearchnet,gu2021quecos,liu2024excs,li2025sacl}, as well as embedding anisotropy and hubness in code representations \citep{li2022competition,gong2023multiview}. 

Graph retrieval offers a promising direction by enabling multi-hop traversal while preserving hierarchical relationships, in contrast to flat index RAG \citep{zhong2024retrieval,wang2023coderag}. By modeling the repository as a graph, where nodes correspond to code entities and edges encode hierarchical or dependency links, GraphRAG can resolve queries that require following transitive dependencies, such as tracing a function call across multiple intermediate layers or modules. However, current graph-based code retrieval methods tend to perform well on code-entity or structure-aware queries, but lack dedicated support for open-ended natural language queries \citep{cao2024repofuse, repograph2025, liu2024graphcoder}.

To address these challenges, we develop an efficient knowledge graph construction procedure together with a Monte Carlo Tree Search (MCTS)-based graph traversal algorithm. Using an agentic architecture, we integrate the knowledge graph with MCTS to enable a dual-stage retrieval system capable of handling both symbolic code-entity queries and natural language queries. Our key contributions are as follows: 

\begin{itemize}[leftmargin=*]
    \item \textbf{Efficient Knowledge Graph Construction for Code Retrieval:} A tool to transform Python repositories into an information-rich knowledge graph that captures hierarchical and cross-file dependencies by parsing abstract syntax trees (AST). To mitigate the semantic gap between natural and symbolic coding languages, we augment graph nodes with textual descriptions of code entities and their corresponding embeddings. 

    \item \textbf{Monte Carlo Tree Search-Based Graph Traversal Algorithm:} A graph traversal algorithm inspired by Monte Carlo Tree Search that balances exploration and exploitation. Starting from a source node, it quickly expands to promising candidates using a bi-encoder. During the simulation phase, a cross-encoder computes reward scores for visited nodes. Over time, rollouts uncover the most relevant node for retrieval. 

    \item \textbf{Router Retrieval Agent:} A dual-stage retrieval pipeline that routes queries by type. Code-entity queries are resolved through fast Cypher lookups on the graph database, while natural language queries fall back to the MCTS-based graph traversal algorithm. 

\end{itemize}

%% file: final_sections/related_work.tex
\paragraph{Code LLMs and Retrieval-Augmented Generation} 
Early neural models for source code established that structure-aware encoders using Abstract Syntax Tree (AST) paths (e.g., code2vec \citep{alon2019code2vec}, code2seq \citep{alon2019code2seq}) or graph neural networks \citep{mou2016tbcnn} \citep{allamanis2018learning} could outperform lexical approaches. Subsequently, Transformer-based pretraining became the dominant paradigm, with models like Codex \citep{chen2021evaluating}, CodeGen \citep{nijkamp2022codegen}, CodeLlama \citep{roziere2023code}, StarCoder2 \citep{lozhkov2024starcoder}, and DeepSeek-Coder \citep{guo2024deepseek} demonstrating strong performance on function- and file-level tasks. However, these models condition on local context and struggle to incorporate the cross-file dependencies essential for reasoning in large repositories.

Early retrieval-augmented generation (RAG) systems such as RECODE \citep{wang2023recode}, REDCODER \citep{parvez2021retrieval}, and TreeGen \citep{sun2020treegen} injected external code snippets into prompts. These methods treated code as flat text, relying on lexical or vector similarity, which hindered their ability to reason across multiple files. While later work improved recall, it remained snippet-centric and failed to model the typed, multi-hop relationships that connect definitions and uses across a codebase.

\paragraph{Natural Language Code Search} 
Natural language--based code search has been extensively studied, beginning with large-scale benchmarks such as CodeSearchNet \citep{husain2019codesearchnet}, which enabled systematic evaluation of neural retrieval models. Subsequent work enriched code embeddings with structural signals, including program dependency graphs  \citealp{wang2020ccgraph}, \citep{chen2024enhancing} and variable flow graphs (deGraphCS, \citealp{zhang2021degraphcs}), while efficiency-focused methods like ExCS \citep{zhang2024excs} improved scalability through offline code expansion. More recently, repository-level approaches employ multi-stage pipelines that integrate commit metadata with BERT re-rankers \citep{sun2025repo} or translate natural language queries into domain-specific query languages \citep{liu2025structural}. In parallel, query reformulation \citep{ye2018queryreform} and LLM-driven paraphrasing\citep{wang2023enhancing} highlight the central challenge of aligning vague natural descriptions with precise code identifiers, especially in large and evolving repositories.
\paragraph{Graph-Based Retrieval and Agentic Frameworks} 
Graph-centric methods address structural limitations by explicitly encoding relationships like definitions, references, and calls, but they differ significantly in scope, persistence, and query support. Some approaches build local graphs, for instance, GraphCoder \citep{liu2024graphcoder} creates Code Context Graphs for snippets but omits cross-file links. CatCoder \citep{catcoder2025} constructs on-the-fly type-dependency subgraphs for statically-typed languages, sacrificing the persistent, long-range relationships needed at repository scale.

Repository-scale graphs improve coverage but introduce trade-offs. RepoGraph \citep{repograph2025} separates definitions and references into distinct nodes with basic invoke/contain edges, which creates redundancy and lacks semantic embeddings for text-code alignment. CoCoMIC \citep{ding-etal-2024-cocomic} models cross-file relations at the file level through imports rather than direct function-to-function edges, constraining multi-hop precision. RepoFuse \citep{cao2024repofuse} uses Jedi-based analysis to build an in-memory graph of imports, inheritance, and calls but focuses on rule-based neighbor capture for completion. Similarly, DraCo \citep{zhang-etal-2024-draco} constructs a fine-grained, variable-level dataflow graph with typed edges (\texttt{Assigns}, \texttt{Refers}, \texttt{Typeof}) but remains specialized for code completion tasks. CodeGraphModel \citep{tao2025codegraphmodel} integrates a repository graph into an LLM via a graph-adapter but relies on lightweight analysis and a simple retrieval method based on entity extraction and string matching, limiting its support for non-entity and multi-hop queries.

A growing line of work couples LLMs with code graphs in agentic frameworks. LocAgent \citep{locagent2025} converts entire codebases into directed graphs and exposes tools like \texttt{SearchEntity} and \texttt{TraverseGraph}, but its comprehensive traversals can be computationally expensive without a persistent graph database. OrcaLoca \citep{orcaloca2025} uses priority-based scheduling and in-memory NetworkX graphs derived from ASTs but acknowledges that its incomplete reference analysis can miss semantic dependencies. CodexGraph \citep{liu2024codexgraphbridginglargelanguage} bridges LLM agents with graph databases for structure-aware retrieval, but its workflows often rely on explicit identifiers, making purely natural language queries challenging. MCTS-based agents like LingmaAgent \citep{ma2024alibaba} explore code graphs with LLM-based reward estimation, while related variants such as RTSoG \citep{long2025enhancing} and REKG-MCTS \citep{zhang-etal-2025-rekg} apply similar strategies to document and text knowledge graphs, but the repeated high-fidelity LLM scoring incurs significant inference cost and can introduce nondeterminism. These trends highlight a need for agents that combine persistent, semantically augmented graphs with cost-aware planning to balance accuracy and efficiency.

This work presents \textbf{RANGER}, a repository-level retrieval agent that integrates persistent graph construction with query-type--aware retrieval. A repository-wide knowledge graph is built through AST parsing and enriched with semantic descriptions and embeddings. At query time, RANGER first converts the input into a Cypher query over this graph. For \emph{code-entity queries}, these Cypher lookups typically suffice for direct resolution. For \emph{natural language queries}, which often fail to return direct matches, RANGER invokes an MCTS-based graph exploration that combines bi-encoder expansion with selective cross-encoder scoring. This dual-path design enables efficient handling of both symbolic and natural language queries, overcoming the limitations of flat embedding indices and gaps of prior graph-based retrieval methods.

%% file: final_sections/methodology.tex
\subsection{Overall Architecture}
We propose a retrieval agent capable of processing both \textit{natural language} and \textit{code-entity} queries for code retrieval. As mentioned earlier, natural language queries are challenging due to the semantic gap between textual descriptions and code embeddings \citep{gu2021codebert, husain2019codesearchnet}.

As illustrated in Figure~\ref{fig:overall}, the system uses a two-stage pipeline with an \emph{offline indexing} stage for repository preprocessing and graph construction and an \emph{online query} stage for retrieval and reasoning with RANGER. In the offline stage, a code repository is parsed into an {entity graph} stored in a graph database (e.g., Neo4j). This includes AST parsing to build the knowledge graph, LLM-assisted description generation for components and modules, and embedding computation for those descriptions.

In the online stage, RANGER first converts the user query into a Cypher statement via zero-shot LLM prompting (prompt in the Appendix). The Cypher query retrieves relevant code entities from the graph database. For \textit{code-entity queries}, these results typically suffice for direct response generation (Path~1). In contrast, \textit{natural language queries} often do not match directly and return \texttt{None}. In such cases, the agent follows {Path 2}, invoking a Monte Carlo Tree Search (MCTS) based graph exploration to iteratively localize the most relevant code snippets. This dual-path design allows RANGER to handle both query types robustly. The following subsections detail the components of this architecture.

\subsection{Code Parsing and Knowledge Graph Creation}\label{kg_creation}

The repository-level knowledge graph is constructed through a two-stage process that first builds isolated file-level graphs and then stitches them into a unified repository-level graph. This design ensures that intra-file structures are captured accurately before resolving complex inter-file dependencies. An illustrative example of this process, using the two-file repository from Figure~\ref{fig:overall}, is provided in Section~\ref{sec:appendix_graph}.

\paragraph{Stage 1: File-level parsing.} Each file is processed using the \texttt{tree-sitter} library \citep{brunsfeld2013tree-sitter}, which produces a detailed Abstract Syntax Tree (AST). This contrasts with existing systems \citep{cao2024repofuse, liu2024codexgraphbridginglargelanguage} that rely on Python-specific tools like Jedi or Parso. We traverse the AST to extract key code entities and relationships, which are organized into an intermediate JSON object serving as a decoupled transfer representation. A database-specific ingestion component then converts these objects into nodes and edges in the graph database. This separation allows new programming languages to be supported by modifying only the AST parser, and new graph backends by updating only the ingestion module. The node types include \texttt{Module}, \texttt{Class}, \texttt{Function}, \texttt{Method}, \texttt{Field}, and \texttt{GlobalVariable}, offering finer granularity than related approaches \citep{ma2024alibaba, locagent2025}. Within each file, structural edges are immediately established, including \texttt{CONTAINS} edges from a \texttt{Module} to its classes and functions, \texttt{HAS\_METHOD} edges from a \texttt{Class} to its methods, and \texttt{INHERITS} edges to represent class inheritance. To handle unresolved dependencies, temporary \texttt{Import} nodes are created, pointing to entities outside the current file. Unlike existing approaches such as the Code Graph Model \citep{tao2025codegraphmodel}, which applies lightweight semantic analysis, or OrcaLoca \citep{orcaloca2025}, which omits static analysis, this step explicitly preserves placeholders for cross-file references.

\paragraph{Stage 2: Repository-level consolidation.} After all files are parsed, the system resolves the temporary \texttt{Import} nodes. Each \texttt{Import} node is matched to its corresponding entity (\texttt{Class}, \texttt{Function}, \texttt{Module}, etc.) elsewhere in the repository, and all incoming edges are redirected to the resolved node. This ``stitching'' step ensures that cross-file dependencies are explicitly represented, yielding broader coverage than prior approaches such as the lightweight cross-file analyses in the Code Graph Model \citep{tao2025codegraphmodel} or the limited function-call tracking in Lingma Agent \citep{ma2024alibaba}. Once redirected, redundant \texttt{Import} nodes are deleted. The result is a repository-level knowledge graph that completely represents both intra-file structure and inter-file dependencies.

\subsection{LLM-Assisted Semantic Description and Embedding}\label{desc_embed}
After constructing the knowledge graph, we add semantic attributes by generating natural language descriptions for each code entity with an LLM using a hierarchical bottom up procedure. Following Code2JSON \citep{singhal2025code2json}, each entity receives two descriptions, a high level purpose summary and a granular member level summary of important variables and behaviors. For small entities such as functions and methods, whose source code fits within the context limit of the LLM, we generate both descriptions directly from code, while for larger entities such as modules and large classes we compose them from precomputed member summaries. We then concatenate the two descriptions, encode them into a vector embedding, and store the text and embedding as node attributes. Prompts are in Appendix~\ref{sec:semantic_prompts}.

\subsection{MCTS-Based Graph Traversal Algorithm} \label{sec:MCTS}

\begin{figure}[ht]
\begin{center}
\includegraphics[width=0.7\textwidth]{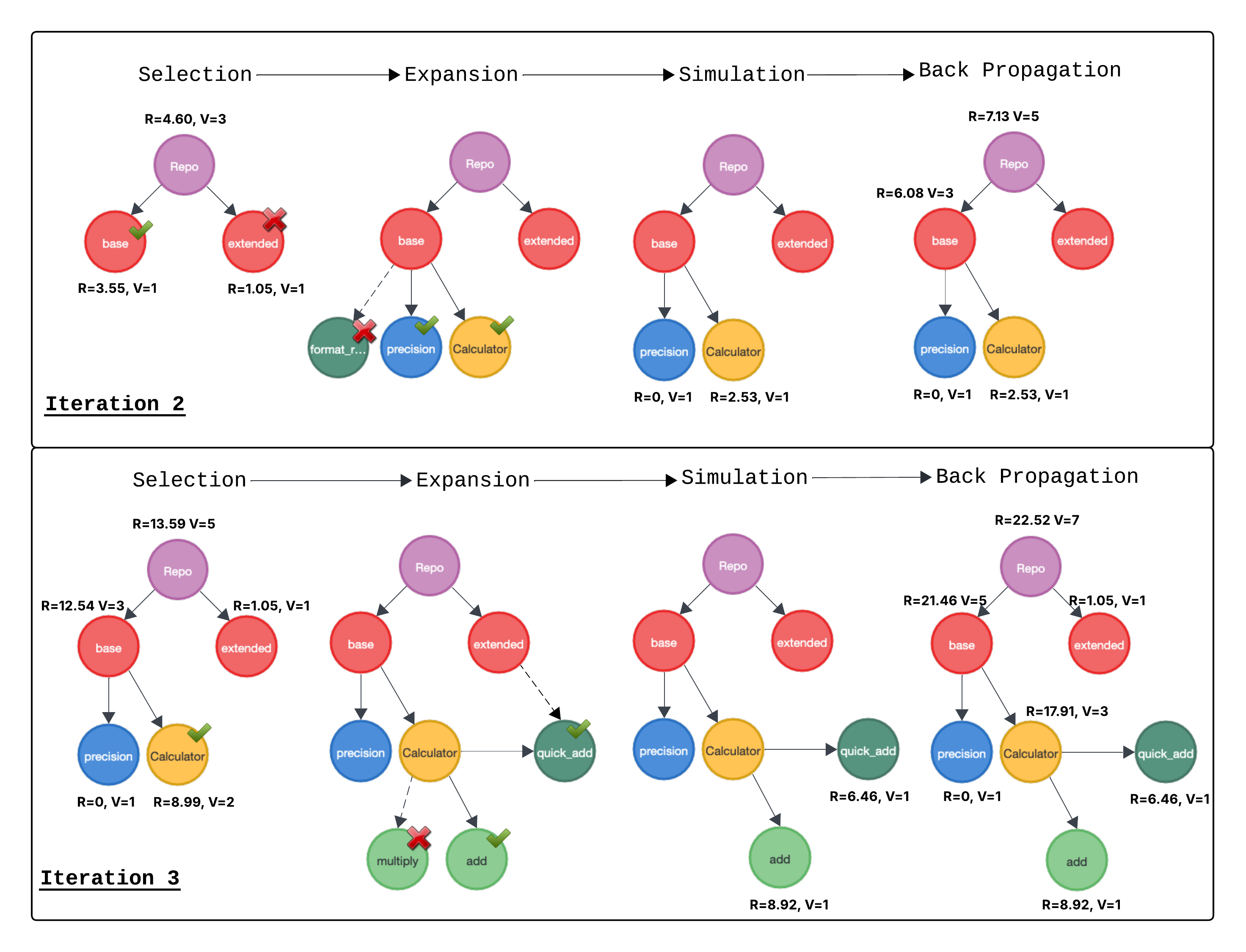}
\end{center}
\caption{
    \textbf{The Monte Carlo Tree Search–based graph traversal algorithm.} 
    The diagram depicts iterations 2 and 3 of a Monte Carlo Tree Search–based 
    graph traversal on the simple two-file code repository knowledge graph 
    from Figure~\ref{fig:overall} in response to the query 
    “Where is the code for addition?” Iteration 2 expands the \texttt{base} 
    module (adding \texttt{precision} and \texttt{Calculator}), simulates their 
    rewards, and back-propagates values to update selection scores. Iteration 3 
    then adds the method node \texttt{add} and function node \texttt{quick\_add}, 
    both of which yield high rewards and are ultimately selected during the 
    extraction and ranking phase as the answer to the user’s query.
}
\label{fig:MCTS}
\end{figure}

To efficiently search the code knowledge graph, we use Monte Carlo Tree Search (MCTS) to balance retrieval efficiency and accuracy. A bi-encoder guides exploration and a cross-encoder scores only the most promising candidates, which focuses computation where expected relevance is highest \citep{wu2019scalable}. The process, formalized in Algorithm~\ref{alg:mcts-graphrag}, consists of Selection, Expansion, Simulation, Backpropagation, and a final Extraction stage.\\

\textbf{Selection. }The selection phase balances exploration (searching new parts of the graph) with exploitation (focusing on paths that have previously yielded high rewards). Starting from the root of the search tree, we recursively select the child node with the highest Upper Confidence bound for Trees (UCT) score, defined as:
\(
\displaystyle
\mathrm{UCT}(v) = \frac{R_v}{\max(1,N_v)} + c\sqrt{\frac{2\ln \max(1,N_{\mathrm{parent}(v)})}{\max(1,N_v)}}
\)
where $R_v$ is the total reward of a node $v$, $N_v$ is its visit count, and $c$ is an exploration parameter. We continue until a leaf is reached. If that leaf is fully expanded, we backtrack to the nearest ancestor with unexpanded neighbors.\\

\textbf{Expansion. }Once a leaf node is selected, the search tree is expanded by adding its neighbors from the code graph as child nodes. To guide this expansion, the bi encoder ranks all neighbors based on the cosine similarity of their embeddings with the query embedding. The top-$k$ most similar and previously unvisited neighbors are then added to the search tree. This bi-encoder driven expansion serves as a fast and effective heuristic for candidate generation.\\

\textbf{Simulation. }This stage evaluates the relevance of newly expanded nodes. Unlike MCTS in adversarial games \citep{silver2017mastering}, where rollouts simulate sequences of actions to a terminal state, our retrieval task lacks a discrete win/loss outcome. A random traversal from a node is ill-suited for determining its relevance to a query. Therefore, we redefine the simulation step as a direct relevance evaluation using a cross-encoder. The query and the node's semantic content are used as input to the cross-encoder, which produces a precise relevance score. This score serves directly as the reward for the node. To maximize throughput, evaluations are processed in batches.\\

\textbf{Backpropagation. }
After evaluation we propagate the reward up the tree. For every node on the path to the root we increment its visit count ($N_v$) and add the reward to its total ($R_v$). This update guides subsequent selection toward promising regions of the code graph.\\

\textbf{Extraction }
After a predefined number of iterations the search terminates and we extract a ranked list of relevant code nodes. The final score for each visited node is
\(
\displaystyle
s(v) = \alpha \cdot \frac{R_v}{\max(1,N_v)} + (1-\alpha)\cdot \mathrm{sim}(E_q, E_v)
\)
which balances the learned MCTS reward with the initial bi encoder similarity to yield a robust final ranking.

%% file: pratik_tables/repobench_result.tex
\begin{table}[h]
\centering
\setlength{\tabcolsep}{6pt}
\renewcommand{\arraystretch}{1.15}
\caption{
Performance comparison on the \textbf{RepoBench} benchmark for cross-file dependency retrieval.
}
\resizebox{\textwidth}{!}{%
\begin{tabular}{l
                S[table-format=1.4]S[table-format=1.4]
                S[table-format=1.4]S[table-format=1.4]
                S[table-format=1.4]S[table-format=1.4]}
\toprule
\multirow{2}{*}{\textbf{Reranker Model}} & \multicolumn{2}{c}{\textbf{Accuracy@5}} & \multicolumn{2}{c}{\textbf{NDCG@5}} & \multicolumn{2}{c}{\textbf{MRR@5}} \\
\cmidrule(lr){2-3} \cmidrule(lr){4-5} \cmidrule(lr){6-7}
 & \textbf{RANGER} & \textbf{Baseline} & \textbf{RANGER} & \textbf{Baseline} & \textbf{RANGER} & \textbf{Baseline} \\
\midrule
Unixcoder-base (110M) & \textbf{0.5446} & 0.4346 & 0.4120 & 0.3075 & 0.3601 & 0.2509 \\
Qwen-3-8B (8B)        & \textbf{0.5471} & 0.4940 & 0.4120 & 0.3530 & 0.3577 & 0.2919 \\

\bottomrule
\end{tabular}
}

\label{tab:repobench_result}
\end{table}

%% file: pratik_tables/crosscodeeval_result.tex
\begin{table}[H]
\centering
\setlength{\tabcolsep}{8pt} 
\renewcommand{\arraystretch}{1.15}

\caption{
Performance comparison of retrieval methods on the \textbf{CrossCodeEval} benchmark for Python.
}

\begin{tabular}{l
                S[table-format=2.2]S[table-format=2.2]
                S[table-format=2.2]S[table-format=2.2]
                S[table-format=2.2]S[table-format=2.2]}
\toprule
\multirow{2}{*}{\textbf{Retrieval Method}} & \multicolumn{2}{c}{\textbf{DeepSeek-Coder-7B}} & \multicolumn{2}{c}{\textbf{CodeLlama-7B}} & \multicolumn{2}{c}{\textbf{StarCoder-7B}} \\
\cmidrule(lr){2-3} \cmidrule(lr){4-5} \cmidrule(lr){6-7}
 & {\textbf{EM}} & {\textbf{ES}} & {\textbf{EM}} & {\textbf{ES}} & {\textbf{EM}} & {\textbf{ES}} \\
\midrule
\textbf{RANGER + BM25} & \textbf{36.27} & 70.77          & \textbf{31.68} & 66.91          & {30.80} & {66.03} \\
BM25                   & 28.57          & 65.95          & 24.87          & 62.83          & {22.33} & {69.60} \\
CGM-MULTI              & 33.88          & 71.19          & {31.03}        & \textbf{73.90} & {\textbf{31.00}} & {71.66} \\
RepoFuse               & 27.92          & 73.09          & 24.80          & 71.05          & {24.20} & {70.82} \\
RLCoder                & 30.28          & \textbf{74.42} & 26.60          & 72.27          & {25.82} & {\textbf{72.11}} \\
R2C2                   & 32.70          & 54.00          & 23.60          & 42.90          & {30.90} & {51.90} \\
\bottomrule
\end{tabular}

\label{tab:retrieval_benchmarking_crosscodeeval}
\end{table}

%% file: final_sections/conclusion.tex
We introduced \textsc{RANGER}, a repository level agent for graph enhanced code retrieval that handles both \textit{code entity queries} and \textit{natural language queries}. This capability is largely absent from existing code retrieval methods. Our MCTS based graph exploration algorithm, most helpful for natural language queries, uses a bi-encoder for expansion and a cross encoder as the reward. On CodeSearchNet and RepoQA we surpass strong semantic retrieval systems, including {Qwen-3-8B} embedding baseline \citep{wang2025qwen3} ranked number one on MTEB Leaderboard \citep{muennighoff2022mteb}, while using smaller models for embedding and reranking \texttt{mxbai-embed-large-v1} with 335M parameters and \texttt{bge-reranker-v2-m3} with 568M parameters. Because cross encoders are more accurate but expensive and often infeasible to apply over the enitre repository, MCTS scores only promising nodes, keeping quality close to exhaustive reranking at lower cost. For repository level completion, where relevant code often lives in other files and is not semantically similar to the query, our graph-guided traversal retrieves the necessary context by following structural relationships rather than embedding proximity alone.

Although \textsc{RANGER} shows strong retrieval performance across multiple benchmarks, several limitations remain. The use of static offline repository graphs limits applicability to dynamic or rapidly evolving codebases where dependencies change frequently. The MCTS stage, while effective for natural language queries, introduces additional inference latency and computational cost that may hinder interactive developer workflows. Node scoring currently depends on cross encoder relevance estimates, which may not be the best reward signal.

Future work will focus on adaptability, efficiency, and evaluation breadth. One direction is incremental graph maintenance that supports live repository updates with minimal recomputation. Another direction is a multi stage retrieval agent in the ReACT style that can combine symbolic Cypher queries with targeted MCTS starting from intermediate graph nodes. This can reduce rollout depth and latency. Learned reward models, including a small language model trained for relevance scoring or reinforcement learning approaches, may offer more robust signals than a fixed cross encoder. At present \textsc{RANGER} supports Python repositories. Since we use the \texttt{tree-sitter} library, which is not Python specific and supports many languages, we plan to extend the system to additional languages. {Code and resources will be released publicly upon acceptance.}

%% file: final_sections/appendix.tex
\subsection {Psuedocode of MCTS algorithm}
Below we present the pesudocode for our Monte Carlo Tree Search - based graph traversal algorithm as described in ~\ref{sec:MCTS} 
\begin{algorithm}[h]
\caption{MCTS-based Graph Traversal Algorithm}
\label{alg:mcts-graphrag}
\begin{algorithmic}[1]
\Require 
\hspace{\algorithmicindent} $q$ \textnormal{(query)}, \quad 
Code graph $\mathcal{G}=(\mathcal{V},\mathcal{E})$ where each $u \in \mathcal{V}$ has description $D_u$ and embedding $E_u=f_{\theta}(u)$, \
\hspace{\algorithmicindent} $r\in\mathcal{V}$ \textnormal{(root repository node)}, \quad
$g_{\phi}:\mathcal{Q}\times\mathcal{D}\to\mathbb{R}$ \textnormal{(cross-encoder)}, \
\hspace{\algorithmicindent} $k_{\text{init}}, k_{\text{min}}\in\mathbb{N}$ \textnormal{(initial \& min expansion width)}, \quad
$c>0$ \textnormal{(UCT exploration)}, \quad
$\alpha\in[0,1]$ \textnormal{(score weighting)}, \
\hspace{\algorithmicindent} $B\in\mathbb{N}$ \textnormal{(retrieval budget)}, \quad
$T\in\mathbb{N}$ \textnormal{(iterations)}
\Ensure Ranked node set $\text{TopK}(\mathcal{V}_{\text{vis}}, B)$ ordered by retrieval score
\Statex \textbf{Notation:} For tree node $v$: visits $N_v$, total reward $R_v$, simulation reward $R_v^{(s)}$, simulation visits $N_v^{(s)}$; 
\[
\text{UCT}(v)=\frac{R_v}{\max(1,N_v)} + c\sqrt{\frac{2\ln \max(1,N_{\text{parent}(v)})}{\max(1,N_v)}}.
\]
Similarity $\text{sim}(E_x,E_y)$ denotes cosine similarity between embeddings.
\State Initialize search tree $\mathcal{T}$ with root $r$; set $N_v\gets0,\, R_v\gets0,\, R_v^{(s)}\gets0,\, N_v^{(s)}\gets0$ for all $v\in\mathcal{T}$
\State Set expansion width $k \gets k_{\text{init}}$; initialize visited nodes $\mathcal{V}_{\text{tree}} \gets \{r\}$
\For{$t=1$ to $T$}
  \Statex \textbf{(A) Selection via UCT}
  \State $\text{curr} \gets r$
  \While{$\text{curr}$ has children in $\mathcal{T}$ and not fully expanded}
    \State $\text{curr} \gets \arg\max_{u \in \text{Children}_{\mathcal{T}}(\text{curr})} \text{UCT}(u)$
  \EndWhile
  \If{$\text{curr}$ is over-visited leaf ($N_{\text{curr}} \geq 2$ and no expandable neighbors)}
    \State Traverse up to find expandable ancestor; if none exists, continue to next iteration
  \EndIf
  \Statex \textbf{(B) Expansion}
  \State $\mathcal{C} \gets \text{Neighbors}_{\mathcal{G}}(\text{curr}) \setminus \mathcal{V}_{\text{tree}}$ \Comment{Non-duplicate children}
  \State $\mathcal{S} \gets \{(u, \text{sim}(E_q, E_u)): u \in \mathcal{C}, E_u \text{ exists}\}$ \Comment{Valid embeddings}
  \If{$\mathcal{S} = \varnothing$}
    \State Mark $\text{curr}$ as fully expanded; \textbf{continue}
  \EndIf
  \State Sort $\mathcal{S}$ by similarity (descending); $\mathcal{E} \gets \text{TopK}(\mathcal{S}, k)$
  \State Add $\mathcal{E}$ as children of $\text{curr}$ in $\mathcal{T}$; $\mathcal{V}_{\text{tree}} \gets \mathcal{V}_{\text{tree}} \cup \mathcal{E}$
  \State Update expansion width: $k \gets \max(k_{\text{min}}, k/2)$ \Comment{Reduce breadth over time}
  \Statex \textbf{(C) Batched Cross-Encoder Simulation}
  \State $\mathcal{P} \gets \{(q, D_u): u \in \mathcal{E}\}$ \Comment{Query-description pairs}
  \State $\mathbf{s} \gets g_{\phi}(\mathcal{P}) \times 10$ \Comment{Batched cross-encoder inference, scale to [0,10]}
  \State $\text{rewards} \gets \{u: \text{clamp}(s_u, 0, 10) \text{ for } u \in \mathcal{E}\}$
  \Statex \textbf{(D) Batched Backpropagation}
  \For{each $(u, r_u) \in \{(u, \text{rewards}[u]): u \in \mathcal{E}\}$}
    \For{each $v$ on path from $u$ to $r$ in $\mathcal{T}$}
      \State $N_v \gets N_v + 1$; $R_v \gets R_v + r_u$
      \State If $v = u$: $R_v^{(s)} \gets R_v^{(s)} + r_u$; $N_v^{(s)} \gets N_v^{(s)} + 1$
    \EndFor
  \EndFor
\EndFor
\Statex \textbf{Final Retrieval Score \& Ranking}
\State $\mathcal{V}_{\text{vis}} \gets \{v \in \mathcal{T}: N_v > 0\}$
\State For each $v \in \mathcal{V}_{\text{vis}}$, compute retrieval score:
\[
s(v) = \alpha \cdot \frac{R_v^{(s)}}{\max(1,N_v^{(s)})} + (1-\alpha) \cdot \text{sim}(E_q, E_v) \times 10
\]
\State \Return $\text{TopK}(\mathcal{V}_{\text{vis}}, B)$ sorted by $s(v)$ (descending)
\end{algorithmic}
\end{algorithm}

\subsection{Knowledge Graph Creation}\label{sec:appendix_graph}
\begin{figure}[h]
\begin{center}

\includegraphics[width=1\textwidth]{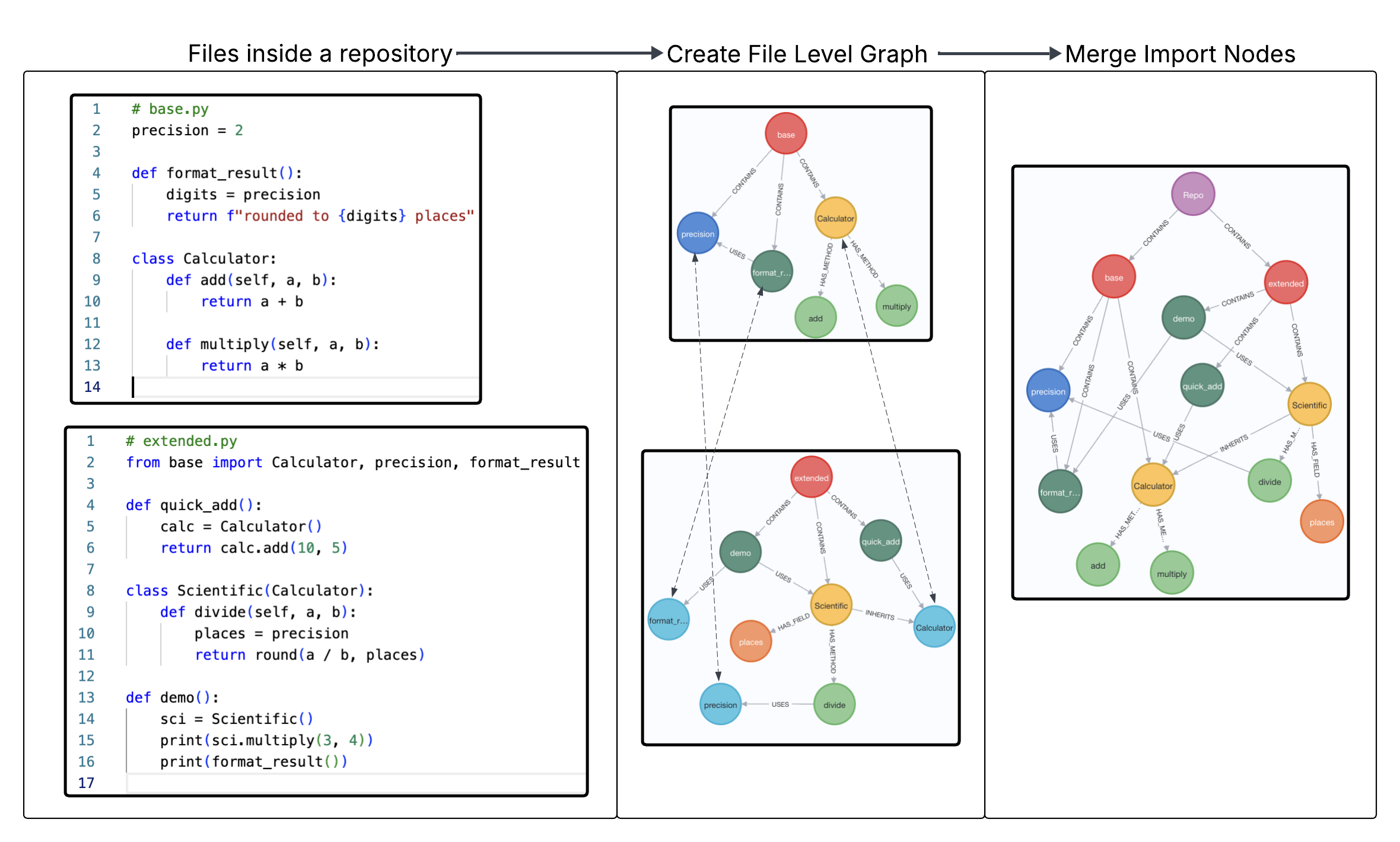}
\end{center}
\caption{
        {The two-stage graph construction process}
    }
\label{fig:graph_indexing}
\end{figure}

Figure~\ref{fig:graph_indexing} illustrates the two-stage repository-level knowledge graph construction process using a simple repository containing \texttt{base.py} and \texttt{extended.py}.

Stage 1 (File-Level Graph Creation): Individual source files undergo Abstract Syntax Tree (AST) parsing using the tree-sitter library to extract granular code entities. For \texttt{base.py}, this creates nodes for the \texttt{base} module, \texttt{Calculator} class, \texttt{add} and \texttt{multiply} methods, \texttt{format\_result} function, and \texttt{precision} global variable. Intra-file hierarchical relationships are established through \texttt{CONTAINS} edges (e.g., \texttt{base} module contains \texttt{Calculator} class and \texttt{format\_result} function), \texttt{HAS\_METHOD} edges (e.g., \texttt{Calculator} class contains \texttt{add} and \texttt{multiply} methods), and \texttt{HAS\_FIELD} edges (e.g., \texttt{base} module contains \texttt{precision} variable).

Similarly, \texttt{extended.py} creates nodes for the \texttt{extended} module, \texttt{Scientific} class, \texttt{divide} method, \texttt{quick\_add} and \texttt{demo} functions. Cross-file dependencies that cannot be immediately resolved are represented as temporary placeholder \texttt{Import} nodes, shown with dashed lines indicating their eventual resolution targets (e.g., imports of \texttt{Calculator}, \texttt{precision}, and \texttt{format\_result} from \texttt{base}).

Stage 2 (Repository-Level Graph Consolidation): The system resolves these \texttt{Import} nodes through multi-step resolution logic, redirecting all incoming relationships to their actual target entities. For example, the \texttt{Scientific} class's inheritance dependency is resolved by establishing a direct \texttt{INHERITS} relationship to the \texttt{Calculator} class, and the \texttt{quick\_add} function's usage of \texttt{Calculator} is connected via a \texttt{USES} relationship. After successful edge redirection, the redundant \texttt{Import} nodes are deleted, resulting in a unified repository-level knowledge graph that captures both hierarchical structure and cross-file dependencies at the variable level.
\subsection {CrossCodeEval Results With  Retrieved Context Limit}\label{sec:cceval_appendix}
\begin{table}[H]
\centering
\setlength{\tabcolsep}{8pt} 
\renewcommand{\arraystretch}{1.15}
\begin{tabular}{l
                S[table-format=2.2]S[table-format=2.2]
                S[table-format=2.2]S[table-format=2.2]
                S[table-format=2.2]S[table-format=2.2]}
\toprule
\multirow{2}{*}{\textbf{Retrieval Method}} & \multicolumn{2}{c}{\textbf{DeepSeek-Coder-7B}} & \multicolumn{2}{c}{\textbf{CodeLlama-7B}} & \multicolumn{2}{c}{\textbf{StarCoder-7B}} \\
\cmidrule(lr){2-3} \cmidrule(lr){4-5} \cmidrule(lr){6-7}
 & {\textbf{EM}} & {\textbf{ES}} & {\textbf{EM}} & {\textbf{ES}} & {\textbf{EM}} & {\textbf{ES}} \\
\midrule
\textbf{RANGER + BM25} & \textbf{34.03} & 69.48          & {29.89} & 66.32          & {26.94} & {64.01} \\
BM25                   & 28.57          & 65.95          & 24.87          & 62.83          & {22.33} & {69.60} \\
CGM-MULTI              & 33.88          & 71.19          & \textbf{31.03}        & \textbf{73.90} & {\textbf{31.00}} & {71.66} \\
RepoFuse               & 27.92          & 73.09          & 24.80          & 71.05          & {24.20} & {70.82} \\
RLCoder                & 30.28          & \textbf{74.42} & 26.60          & 72.27          & {25.82} & {\textbf{72.11}} \\
R2C2                   & 32.70          & 54.00          & 23.60          & 42.90          & {30.90} & {51.90} \\
\bottomrule
\end{tabular}
\caption{
Performance comparison of retrieval methods on the \textbf{CrossCodeEval} benchmark for Python with a limit on retrieved context of 4,096 tokens .
}
\label{tab:retrieval_benchmarking_crosscodeeval_appendix}
\end{table}

\subsection{Experimental parameters}
In this section, we provide the experimental parameters corresponding to the results reported in  Section~\ref{sec:experiments}.
\begin{table}[H]
\centering
\setlength{\tabcolsep}{8pt}
\renewcommand{\arraystretch}{1.15}

\label{tab:exp_params}
\begin{tabular}{ll}
\toprule
\textbf{Parameter} & \textbf{Specification} \\
\midrule

\multicolumn{2}{l}{\textbf{RepoBench }} \\
\quad Cypher Generator  & hugging-quants/Meta-Llama-3.1-70B-Instruct-AWQ-INT4 \\
\midrule
\multicolumn{2}{l}{\textbf{CrossCodeEval }} \\
\quad Cypher Generator  & hugging-quants/Meta-Llama-3.1-70B-Instruct-AWQ-INT4 \\
\midrule
\multicolumn{2}{l}{\textbf{CodeSearchNet }}  \\
\quad Query Embedding (MCTS) & mxbai-embed-large-v1 (335 M Params) \\
\quad Text Description Generation & deepseek-coder-1.3B-instruct \\
\quad MCTS Cross-Encoder & bge-reranker-v2-m3 (568 M Params)\\
\quad MCTS max number of iterations &200\\
\quad Total number of `Module' Nodes in graph & $N$\\
\quad MCTS $k_{init}$ & $N//2$ \\
\quad MCTS $k_{min}$ & 20\\
\quad MCTS $c$ (exploration constant) & $\frac{1}{(8\sqrt{\ln({2*N})})}$\\
\quad MCTS $\alpha$ & 0.5\\

\midrule
\multicolumn{2}{l}{\textbf{RepoQA }} \\
\quad Query Embedding (MCTS) & mxbai-embed-large-v1 (335 M Params) \\
\quad Text Description Generation & deepseek-coder-1.3B-instruct \\
\quad MCTS Cross-Encoder & bge-reranker-v2-m3 (568 M Params)\\
\quad MCTS max number of iterations &500\\
\quad Total number of `Module' Nodes in graph & $N$\\
\quad MCTS $k_{init}$ & $N//2$ \\
\quad MCTS $k_{min}$ & 20\\
\quad MCTS $c$ (exploration constant) & $\frac{1}{(\sqrt{\ln({4*N})})}$\\
\quad MCTS $\alpha$ & 0.9\\

\bottomrule
\end{tabular}
\caption{
Experimental parameters for batch/online processing and evaluation benchmarks.
}
\end{table}
\subsection{Examples of Generated Cypher Queries}

In this section, we provide examples of Cypher queries generated for different code completion and search tasks across various datasets. Tables ~\ref{tab:codesearchnet_llm_cypher_rows} and ~\ref{tab:repobench_cypher_example}  demonstrate the query generation process including entity identification, model reasoning, and the resulting Cypher queries.
\begin{table}[htbp]
\centering
\footnotesize
\caption{Natural language query and generated Cypher query example from CodeSearchNet dataset}
\label{tab:codesearchnet_llm_cypher_rows}
\begin{tabular}{>{\raggedright}p{3.5cm} >{\raggedright\arraybackslash}p{10.8cm}}
\toprule
\textbf{Natural Language Query} & ``how to get database table name'' \\
\midrule
\textbf{Generated Cypher Query} & 
\begin{lstlisting}[basicstyle=\scriptsize\ttfamily, breaklines=true, frame=none]
MATCH (r:Repo)-[:CONTAINS]->(m:Module {name: 'database'})
      -[:CONTAINS]->(c:Class)
RETURN c.name, c.code
\end{lstlisting} \\
\bottomrule
\end{tabular}
\end{table}

\begin{table}[H]
\centering
\footnotesize
\caption{Entity query, model reasoning, and generated Cypher query for completing the \texttt{test\_renderables} function in RepoBench dataset}
\label{tab:repobench_cypher_example}
\begin{tabular}{>{\raggedright}p{2.8cm} >{\raggedright\arraybackslash}p{11.5cm}}
\toprule
\textbf{Entity Query} & 
\textbf{Input:} \texttt{file\_name: tests.test\_renderables} \\
& Fetch the most important dependencies from the repository to complete the following code: \\[0.2cm]
& \begin{lstlisting}[language=Python, basicstyle=\scriptsize\ttfamily, breaklines=true, frame=none]
# Copyright (C) 2023  ETH Zurich, Manuel Kaufmann, 
# Velko Vechev, Dario Mylonopoulos
@reference()
def test_renderables(viewer: Viewer):
    grid_xz = np.mgrid[-1.5:1.5:0.3, -1.5:1.5:0.3]
    n_lines = grid_xz.shape[1] * grid_xz.shape[2]
    xz_coords = np.reshape(grid_xz, (2, -1)).T
    line_starts = np.concatenate([
        xz_coords[:, 0:1], 
        np.zeros((n_lines, 1)), 
        xz_coords[:, 1:2]
    ], axis=-1)
    line_ends = line_starts.copy()
    line_ends[:, 1] = 1.0
    line_strip = np.zeros((2 * n_lines, 3))
    line_strip[::2] = line_starts
    line_strip[1::2] = line_ends
\end{lstlisting} \\
\midrule
\textbf{LLM Thoughts} & Incomplete element identified: function \texttt{test\_renderables}. \\
\midrule
\textbf{Generated Cypher Query} & 
\begin{lstlisting}[basicstyle=\scriptsize\ttfamily, breaklines=true, frame=none]
MATCH (m:Module {name: 'tests.test_renderables'})
      -[:CONTAINS]->(f:Function {name: 'test_renderables'})
OPTIONAL MATCH (f)-[:USES]->(dep)
RETURN DISTINCT dep.name AS name, 
       dep.signature AS signature, 
       dep.code AS code
\end{lstlisting} \\
\bottomrule
\end{tabular}
\end{table}

\subsection{System Prompt for CrossCodeEval Dataset}

The following system prompt is used for generating Cypher queries in the CrossCodeEval evaluation setup. This prompt guides the model to generate precise queries for cross-file dependency analysis while maintaining proper syntax and semantic correctness.

\begin{lstlisting}[basicstyle=\footnotesize\ttfamily, breaklines=true, frame=single, backgroundcolor=\color{lightgray}]
# Neo4j Cypher Query Expert for Code Dependency Analysis

You are a Neo4j Cypher query expert. Your task is to generate concise 
Cypher queries to find ALL cross file dependencies that will help to 
complete the provided incomplete code based on the provided graph schema.

## Graph Schema:
**Nodes**: Repo (name), Module, Class (name, code, signature, 
module_name), Function (name, code, signature, module_name), 
Method (name, code, signature, module_name, class)
**Edges**: CONTAINS (Repo->Module, Module->Class/Function), 
HAS_METHOD (Class->Method), INHERITS (Class->Class), 
USES (All->Dependencies)

## Instructions:
- **CORRECTNESS**: Use proper Cypher syntax. Ensure each UNION branch 
  in Cypher has a complete MATCH...RETURN with SAME COLUMN NAMES.
- **GENERATE MINIMAL QUERIES**: ONLY RETRIEVE THOSE NODES THAT YOU 
  WILL REQUIRE TO COMPLETE THE INCOMPLETE CODE. Use fewest UNION 
  clauses possible.
- **MANDATORY**: Return the entire nodes as ***dep*** and their labels 
  as ***label*** in the query. NOTE THE NAMES SHOULD BE 'dep' and 
  'label' ONLY.
- **IMPORTANT**: PAY EXTRA ATTENTION TO THE LAST INCOMPLETE LINE, THE 
  FUNCTION/METHOD/CLASS BEING USED IN THE LAST INCOMPLETE LINE, AND 
  TRACE THEM TO WHERE THEY ARE INSTANTIATED/IMPORTED, TO FETCH 
  CORRECT DEPENDENCIES.
- **IMPORTANT**: PAY EXTRA ATTENTION TO IMPORT ALIASES, AND ONLY THE 
  GLOBAL VARIABLES BEING USED IN THE LAST INCOMPLETE LINE.
- **IMPORTANT**: In the generated cypher query ONLY USE NAMES YOU ARE 
  CONFIDENT ABOUT OR ELSE DON'T USE THEM. For imports, avoid module 
  names as they may differ. It is fine if we get some false positives.
- **IMPORTANT**: PAY ATTENTION TO THE PROVIDED GRAPH SCHEMA TO MAKE 
  CORRECT QUERIES.

## Input Data Format:
Given repo_name: Repository name which can use to identify the Repo 
Node in the graph.
Given file_name: File name which can use to identify the Module Node 
in the graph.
Fetch the most important connected nodes from the graph to predict the 
next line of the below code:
Incomplete code snippet to complete.

## Your Task:
First provide a brief thought on your decision process, then generate 
**ONLY THE CYPHER QUERY**.

**Format:**
```
**Thought:** Incomplete element identified: <element_name> 
(function/method)
**Query:**
[Cypher query only]
```

## Example
Given repo_name: /Users/pratik.shah1/work/CrossCodeEval_repos/
google_alert-system
Given file_name: models.classes
Fetch the most important connected nodes from the graph to predict the 
next line of the below code:

import numpy as np
from poptransformer import ops
from poptransformer.layers.layer_norm import BaseLayerNorm
from classes import BaseModule as base_module

class BaseRMSLayerNorm(BaseLayerNorm):
    def __init__(self, input_size, eps=1e-5, context=''):
        self.base_object = base_module()
        
    def collect_bind_layer_weights(self):
        weight_key = '.'.join([self.context, 'weight'])
        weight_np = self.get_param_from_state_dict(weight_key, 
                                                  [self.input_size])
        self.weight_id = self.add_initialized_input_tensor(weight_np, 
                                                          weight_key)
        
    def __call__(self, graph, x):
        variance_epsilon = ops.constant(graph, 
                                      np.array(self.eps).astype(np.float32), 
                                      'variance_epsilon')
        variance = self.base_object.

**Thought:** Incomplete method __call__ in BaseRMSLayerNorm class, 
remaining methods are not important. The last incomplete line uses 
self.base_object, which calls base_module but that is an ALIAS of the 
imported BaseModule class suggesting need for BaseModule. Also need 
parent class BaseLayerNorm for inheritance context.

**Query:**
```cypher
MATCH (r:Repo {name: '/Users/pratik.shah1/work/CrossCodeEval_repos/
google_alert-system'})-[:CONTAINS]->(m:Module)-[:CONTAINS]->
(c:Class {name: 'BaseRMSLayerNorm'})-[:HAS_METHOD]->
(method {name: '__call__'})-[:USES]->(dep)
RETURN DISTINCT dep, labels(dep) as label
UNION
MATCH (r:Repo {name: '/Users/pratik.shah1/work/CrossCodeEval_repos/
google_alert-system'})-[:CONTAINS]->(m:Module)-[:CONTAINS]->
(c:Class {name: 'BaseModule'})
RETURN DISTINCT c as dep, labels(c) as label
UNION
MATCH (r:Repo {name: '/Users/pratik.shah1/work/CrossCodeEval_repos/
google_alert-system'})-[:CONTAINS]->(m:Module)-[:CONTAINS]->
(c:Class {name: 'BaseLayerNorm'})
RETURN DISTINCT c as dep, labels(c) as label
```
\end{lstlisting}

\subsection{System Prompt for RepoBench Dataset}

The RepoBench system prompt is specifically designed for repository-level code completion tasks, with enhanced decision-making logic for identifying incomplete code elements and generating appropriate Cypher queries.

\begin{lstlisting}[basicstyle=\footnotesize\ttfamily, breaklines=true, frame=single, backgroundcolor=\color{lightgray}]
# Neo4j Cypher Query Expert for Code Dependency Analysis

You are a Neo4j Cypher query expert. Your task is to generate concise 
Cypher queries to find dependencies for code snippets based on the 
provided graph schema.

## Decision Process:
1. **ANALYZE CODE COMPLETENESS**: Check if there's an incomplete element 
   near the bottom of the code snippet
2. **IF COMPLETE**: Use global fallback approach for file-level 
   dependencies  
3. **TO FIND INCOMPLETE**: 
   - 3a **For collections/lists/dicts**: Missing closing bracket `]`, 
       `}`, or `)`
   - 3b **For functions/classes/methods**: Missing body, incomplete 
       signature, or abrupt termination
4. **CRITICAL**: Only use visible information. DO NOT GUESS incomplete 
   elements if their definitions aren't clearly shown. **NEVER ASSUME** 
   - if unsure, always use global fallback.
   - **ONLY** identify incomplete elements if you see actual `def`, 
     `class`, or variable assignment with collections `[`, `{`
   - Don't identify based on function calls/usage or comments

## Instructions:
- **CRITICAL**: When you find an incomplete function/method/class/
  collection, you MUST identify its name and use the specific template 
  for that element - BUT ONLY if the definition is clearly visible
- **NO GUESSING**: If the element definition is not clearly shown, use 
  global fallback instead
- **MUST SEE**: Actual `def function():`, `class Name:`, or 
  `variable = [` syntax to identify incomplete elements
- **INDENTATION MATTERS**: Pay close attention to indentation to 
  distinguish functions (no indent) vs methods (indented) - this is 
  crucial for correct queries
- **GENERATE MINIMAL QUERIES**: Use fewest UNION clauses possible
- **MANDATORY**: Return ONLY 'name', 'code', 'signature' attributes
- **IMPORTANT**: Pay attention to complete file path including folder names
- **MODULE NODES**: Use 'name' for dotted names, 'local_name' for 
  undotted names
- **CORRECTNESS**: Use proper Cypher syntax. Ensure each UNION branch 
  in Cypher has a complete MATCH...RETURN with identical column names 
  and orders

## Graph Schema:
**Nodes**: Module (name, local_name, code, signature), 
Class (name, code, signature, module_name), 
Function (name, code, signature, module_name), 
Method (name, code, signature, module_name, class), 
Field (name, class), 
GlobalVariable (name, code, module_name)
**Edges**: CONTAINS (Module->Class/Function/GlobalVariable), 
HAS_METHOD (Class->Method), HAS_FIELD (Class->Field), 
INHERITS (Class->Class), USES (All->Dependencies)

## Example Queries:

### Example 1 - Incomplete Method
**User Query:**
```
Given file_name: src.alert.interference.reporting.admin.admin
Fetch dependencies for code:
    def get_form_class(self, request, obj=None):
        return ColumnTemplateForm(request)
    def get_client_data(self, request):
```

**Thought:** Incomplete element identified: method `get_client_data` 
(based on indentation).

**Query:**
```cypher
MATCH (m:Module {name: 'src.alert.interference.reporting.admin.admin'})
      -[:CONTAINS]->(c:Class {name: 'ColumnTemplateAdmin'})
      -[:HAS_METHOD]->(method {name: 'get_client_data'})
OPTIONAL MATCH (method)-[:USES]->(dep)
RETURN DISTINCT dep.name AS name, 
       dep.signature AS signature, 
       dep.code AS code
```

## Your Task:
First provide a brief thought on your decision process, then generate 
**ONLY THE CYPHER QUERY**.

**Format:**
```
**Thought:** [Incomplete element identified: <element_name> OR 
No incomplete element identified]
**Query:**
[Cypher query only]
```

## User Query:
```
\end{lstlisting}

\subsection{Graph Schema}

\begin{lstlisting}[basicstyle=\footnotesize\ttfamily, breaklines=true, frame=single, backgroundcolor=\color{lightgray}]
# Graph Schema Description

## Nodes and Attributes:

1. **Module**:
   - **Attributes:**
     - `name` (String): Dotted module name
     - `local_name` (String): Local module name (no path)
     - `embedding` (Vector): Embedding from module description
     - `description` (String): Summary of the module

2. **Class**:
   - **Attributes:**
     - `name` (String): Class name
     - `signature` (String): Class signature
     - `code` (String): Full class code
     - `module_name` (String): Owning module name
     - `embedding` (Vector): Embedding from description and member_descriptions
     - `description` (String): High-level summary of the class
     - `member_descriptions` (String): Descriptions of constituent members

3. **Function**:
   - **Attributes:**
     - `name` (String): Function name
     - `code` (String): Full function code
     - `signature` (String): Function signature
     - `module_name` (String): Owning module name
     - `embedding` (Vector): Embedding from description and member_descriptions
     - `description` (String): High-level summary of the function
     - `member_descriptions` (String): Descriptions of constituent elements

4. **Field**:
   - **Attributes:**
     - `name` (String): Field name
     - `code` (String): Definition code segment
     - `class` (String): Owning class name
     - `description` (String): Summary of the field
     - `member_descriptions` (String): Details of field usage
     - `embedding` (Vector): Embedding from description and member_descriptions

5. **Method**:
   - **Attributes:**
     - `name` (String): Method name
     - `class` (String): Owning class name
     - `code` (String): Full method code
     - `signature` (String): Method signature
     - `module_name` (String): Owning module name
     - `embedding` (Vector): Embedding from description and member_descriptions
     - `description` (String): High-level summary of the method
     - `member_descriptions` (String): Descriptions of method members

6. **GlobalVariable**:
   - **Attributes:**
     - `name` (String): Global variable name
     - `code` (String): Definition code segment
     - `module_name` (String): Owning module name
     - `embedding` (Vector): Embedding from description and member_descriptions
     - `description` (String): Summary of the variable
     - `member_descriptions` (String): Details of variable usage

7. **Repo**:
   - **Attributes:**
     - `name` (String): Repository name

8. **Import**: (temporary)
   - **Attributes:**
     - `name` (String): Imported item name
     - `module` (String): Source module name
     - `alias` (String, optional): Alias used in import
     - `dotted_folder_name` (String, optional): Submodule path

## Edges and Relationships:

1. **CONTAINS**:
   - **Source:** `Module` or `Repo`
   - **Target:** `Module`, `Class`, `Function`, or `GlobalVariable`

2. **HAS_METHOD**:
   - **Source:** `Class`
   - **Target:** `Method`

3. **HAS_FIELD**:
   - **Source:** `Class`
   - **Target:** `Field`

4. **INHERITS**:
   - **Source:** `Class`
   - **Target:** `Class` (base class)

5. **USES**:
   - **Source:** `Class`, `Function`, `Method`, or `GlobalVariable`
   - **Target:** `Class`, `Function`, `Method`, or `GlobalVariable`
\end{lstlisting}

\subsection{Prompts for generation of Semantic Description of Entities}\label{sec:semantic_prompts}

Below are the three prompt templates used to generate high-level and member-specific descriptions for each code entity, as well as the summarization prompt for larger entities (e.g., summarizing the descriptions of all constituent classes, functions, and variables for modules).

\subsubsection{Code Summarization Prompt}

\begin{lstlisting}[basicstyle=\footnotesize\ttfamily, breaklines=true, frame=single, backgroundcolor=\color{lightgray}]
### Task: Code Summarization

Summarize the code at a high level without referencing specific function 
or variable names. Focus on its purpose, how it is implemented, and its 
notable features. Use the following format:

**PURPOSE**
Describe what the code is designed to achieve.

**IMPLEMENTATION**
Explain how the code accomplishes its purpose, including general 
techniques or components used, without naming exact functions or variables.

**KEY FEATURES**
List significant capabilities, design patterns, or behaviors the code 
exhibits.

### Programming Language: Python
### Code:

\end{lstlisting}

\subsubsection{Code Members Description Prompt}

\begin{lstlisting}[basicstyle=\footnotesize\ttfamily, breaklines=true, frame=single, backgroundcolor=\color{lightgray}]
### Task: Code Members Description

Analyze the Python code and identify important variables (skip temporary 
variables and trivial assignments), functions and classes (also function 
calls and class instantiations). Use the following format:

name - description

List each important code member with its name followed by a dash and a 
*** one-line short description *** of its purpose or functionality.

If no important members are found, respond with: ---None---

***DO NOT REPEAT MEMBERS. YOU CAN CONCLUDE EARLY ONCE ALL MEMBERS ARE 
LISTED.***

### Programming Language: Python
### Code:

\end{lstlisting}

\subsubsection{File Summary from Component Descriptions Prompt}

\begin{lstlisting}[basicstyle=\footnotesize\ttfamily, breaklines=true, frame=single, backgroundcolor=\color{lightgray}]
### Task: File Summary from Component Descriptions

Create a high-level summary of a Python file based on the provided component 
descriptions. You are not given any code, but only the descriptions of 
parts of the code given by various developers. You have to use ALL these 
descriptions to summarize the code.

### Guidelines:
1. Do not include any code in your response, or guess the code. Simply 
   try and summarize the descriptions provided to you.
2. Focus on the file's overall purpose, architecture, key functionality, 
   and key members.
3. If no description is provided simply say 'No description found'.
4. Summarize the purpose of ALL components mentioned in the descriptions.

\end{lstlisting}
\subsection{LLM Usage}
We used large language models solely for grammar and style polishing. We are fully accountable for all ideas, analyses, and claims, which were authored and verified by us.

%% file: RANGER-submission.bbl
\begin{thebibliography}{61}
\providecommand{\natexlab}[1]{#1}
\providecommand{\url}[1]{\texttt{#1}}
\expandafter\ifx\csname urlstyle\endcsname\relax
  \providecommand{\doi}[1]{doi: #1}\else
  \providecommand{\doi}{doi: \begingroup \urlstyle{rm}\Url}\fi

\bibitem[Allamanis et~al.(2018{\natexlab{a}})Allamanis, Barr, Devanbu, and Sutton]{allamanis2018survey}
Miltiadis Allamanis, Earl~T Barr, Premkumar Devanbu, and Charles Sutton.
\newblock A survey of machine learning for big code and naturalness.
\newblock \emph{ACM Computing Surveys}, 51\penalty0 (4):\penalty0 1--37, 2018{\natexlab{a}}.

\bibitem[Allamanis et~al.(2018{\natexlab{b}})Allamanis, Brockschmidt, and Khademi]{allamanis2018learning}
Miltiadis Allamanis, Marc Brockschmidt, and Mahmoud Khademi.
\newblock Learning to represent programs with graphs.
\newblock In \emph{ICLR}, 2018{\natexlab{b}}.

\bibitem[Alon et~al.(2019{\natexlab{a}})Alon, Brody, Levy, and Yahav]{alon2019code2seq}
Uri Alon, Shaked Brody, Omer Levy, and Eran Yahav.
\newblock code2seq: Generating sequences from structured representations of code.
\newblock In \emph{ICLR}, 2019{\natexlab{a}}.

\bibitem[Alon et~al.(2019{\natexlab{b}})Alon, Zilberstein, Levy, and Yahav]{alon2019code2vec}
Uri Alon, Meital Zilberstein, Omer Levy, and Eran Yahav.
\newblock code2vec: Learning distributed representations of code.
\newblock In \emph{POPL}, 2019{\natexlab{b}}.

\bibitem[Brunsfeld et~al.(2013)]{brunsfeld2013tree-sitter}
Max Brunsfeld et~al.
\newblock {Tree-sitter: An incremental parsing system for programming tools}.
\newblock \\url{https://github.com/tree-sitter/tree-sitter}, 2013.
\newblock Accessed: 2025-09-16.

\bibitem[Cao et~al.(2024)Cao, Zhen, Fan, and Gao]{cao2024repofuse}
Zixuan Cao, Yuxin Zhen, Gang Fan, and Shuo Gao.
\newblock Repository-level code completion with fused dual context.
\newblock \emph{arXiv preprint arXiv:2402.14323}, 2024.

\bibitem[Chen et~al.(2021)Chen, Tworek, Jun, Yuan, Pinto, Kaplan, Edwards, Burda, Joseph, Brockman, Ray, Puri, Krueger, Petrov, Khlaaf, Sastry, Mishkin, Chan, Gray, Ryder, Pavlov, Power, Kaiser, Bavarian, Winter, Tillet, Such, Cummings, Plappert, Chantzis, Barnes, Herbert-Voss, Guss, Nichol, Paino, Tezak, Tang, Babuschkin, Balaji, Jain, Saunders, Hesse, Carr, Leike, Achiam, Misra, Morikawa, Radford, Sutskever, Amodei, et~al.]{chen2021evaluating}
Mark Chen, Jerry Tworek, Heewoo Jun, Qiming Yuan, Henrique Pinto, Jared Kaplan, Harri Edwards, Yuri Burda, Nicholas Joseph, Greg Brockman, Alex Ray, Gursimar Puri, Gretchen Krueger, Michael Petrov, Heidy Khlaaf, Girish Sastry, Pamela Mishkin, Brooke Chan, Scott Gray, Nick Ryder, Mikhail Pavlov, Barnabas Power, Lukas Kaiser, Mohammad Bavarian, Clemens Winter, Phil Tillet, Felipe~Petroski Such, David Cummings, Matthias Plappert, Fotios Chantzis, Elizabeth Barnes, Ariel Herbert-Voss, William~H Guss, Alex Nichol, Alex Paino, Nikolas Tezak, Jie Tang, Igor Babuschkin, Suchir Balaji, Shantanu Jain, William Saunders, Christopher Hesse, Andrew~N Carr, Jan Leike, Joshua Achiam, Vedant Misra, Emy Morikawa, Alec Radford, Ilya Sutskever, Dario Amodei, et~al.
\newblock Evaluating large language models trained on code.
\newblock \emph{arXiv preprint arXiv:2107.03374}, 2021.

\bibitem[Chen et~al.(2024)Chen, Liu, and Ge]{chen2024enhancing}
Yifan Chen, Yang Liu, and Long Ge.
\newblock Enhancing source code summarization with a hierarchical structural-aware transformer based on program dependency graph.
\newblock \emph{arXiv preprint arXiv:2409.06208}, 2024.

\bibitem[Chen et~al.(2025)Chen, Tang, Deng, Wu, Wu, Jiang, Prasanna, Cohan, and Wang]{locagent2025}
Zhaoling Chen, Xiangru Tang, Gangda Deng, Fang Wu, Jialong Wu, Zhiwei Jiang, Viktor Prasanna, Arman Cohan, and Xingyao Wang.
\newblock Locagent: Graph-guided llm agents for code localization.
\newblock In \emph{Proceedings of the 2025 Annual Meeting of the Association for Computational Linguistics (ACL)}, 2025.

\bibitem[Ding et~al.(2023)Ding, Wang, Ahmad, Ding, Tan, Jain, Ramanathan, Nallapati, Bhatia, Roth, et~al.]{ding2023crosscodeeval}
Yangruibo Ding, Zijian Wang, Wasi Ahmad, Hantian Ding, Ming Tan, Nihal Jain, Murali~Krishna Ramanathan, Ramesh Nallapati, Parminder Bhatia, Dan Roth, et~al.
\newblock Crosscodeeval: A diverse and multilingual benchmark for cross-file code completion.
\newblock \emph{Advances in Neural Information Processing Systems}, 36:\penalty0 46701--46723, 2023.

\bibitem[Ding et~al.(2024)Ding, Wang, Ahmad, Ramanathan, Nallapati, Bhatia, Roth, and Xiang]{ding-etal-2024-cocomic}
Yangruibo Ding, Zijian Wang, Wasi~Uddin Ahmad, Murali~Krishna Ramanathan, Ramesh Nallapati, Parminder Bhatia, Dan Roth, and Bing Xiang.
\newblock {C}o{C}o{MIC}: Code completion by jointly modeling in-file and cross-file context.
\newblock In \emph{Proceedings of the 2024 Joint International Conference on Computational Linguistics, Language Resources and Evaluation (LREC-COLING 2024)}, Torino, Italy, May 2024. ELRA and ICCL.
\newblock URL \url{https://aclanthology.org/2024.lrec-main.305}.

\bibitem[Gong et~al.(2023)Gong, Xu, Chen, and Kim]{gong2023multiview}
Yu~Gong, Bowen Xu, Zheng Chen, and Miryung Kim.
\newblock Multi-view contrastive learning for code search.
\newblock In \emph{Proceedings of the 45th International Conference on Software Engineering}, 2023.

\bibitem[Gu et~al.(2021{\natexlab{a}})Gu, Ren, Lou, Zhang, Liu, Huang, Li, Sun, and Zhou]{gu2021codebert}
Xiaodong Gu, Shan Ren, Shuo Lou, Daniel Zhang, Alex Liu, Wei Huang, Ge~Li, Zhi~Jin Sun, and Michael R~Lyu Zhou.
\newblock Codebert: A pre-trained model for programming and natural languages.
\newblock \emph{Neurocomputing}, 453:\penalty0 293--301, 2021{\natexlab{a}}.

\bibitem[Gu et~al.(2021{\natexlab{b}})Gu, Ren, Zhang, and Kim]{gu2021quecos}
Xiaodong Gu, Shuo Ren, Yao Zhang, and Sunghun Kim.
\newblock Enriching query semantics for code search with reinforcement learning.
\newblock In \emph{Proceedings of the 43rd International Conference on Software Engineering (ICSE)}, pp.\  1550--1561. IEEE, 2021{\natexlab{b}}.
\newblock URL \url{https://arxiv.org/abs/2105.09630}.

\bibitem[Guo et~al.(2022)Guo, Ren, Lu, Feng, Tang, Duan, Zhou, Yin, Shou, Jiang, et~al.]{guo2022unixcoder}
Daya Guo, Shuo Ren, Shuai Lu, Zhangyin Feng, Duyu Tang, Nan Duan, Long Zhou, Jian Yin, Linjun Shou, Daxin Jiang, et~al.
\newblock {UnixCoder}: Unified cross-modal pre-training for code representation.
\newblock In \emph{Proceedings of the 60th Annual Meeting of the Association for Computational Linguistics}, 2022.

\bibitem[Guo et~al.(2024)Guo, Zhu, Yang, Xie, Dong, Zhang, Chen, Bi, Wu, Li, et~al.]{guo2024deepseek}
Daya Guo, Qihao Zhu, Dejian Yang, Zhenda Xie, Kai Dong, Wentao Zhang, Guanting Chen, Xiao Bi, Y~Wu, YK~Li, et~al.
\newblock Deepseek-coder: When the large language model meets programming--the rise of code intelligence.
\newblock \emph{arXiv preprint arXiv:2401.14196}, 2024.

\bibitem[Husain et~al.(2019)Husain, Wu, Gazit, Allamanis, and Brockschmidt]{husain2019codesearchnet}
Hamel Husain, Ho-Hsiang Wu, Tiferet Gazit, Miltiadis Allamanis, and Marc Brockschmidt.
\newblock Codesearchnet challenge: Evaluating the state of semantic code search.
\newblock \emph{arXiv preprint arXiv:1909.09436}, 2019.

\bibitem[Izacard et~al.(2022)Izacard, Petroni, Hosseini, Grave, and Riedel]{izacard2022unsupervised}
Gautier Izacard, Fabio Petroni, Seyed Mehran~Kazemi Hosseini, Edouard Grave, and Sebastian Riedel.
\newblock Unsupervised dense information retrieval with contrastive learning.
\newblock In \emph{Transactions of the Association for Computational Linguistics}, 2022.

\bibitem[Karpukhin et~al.(2020)Karpukhin, Oguz, Min, Wu, Edunov, Chen, and Yih]{karpukhin2020dense}
Vladimir Karpukhin, Barlas Oguz, Sewon Min, Ledell Wu, Sergey Edunov, Danqi Chen, and Wen-tau Yih.
\newblock Dense passage retrieval for open-domain question answering.
\newblock In \emph{Proceedings of the 2020 Conference on Empirical Methods in Natural Language Processing}, 2020.

\bibitem[Li et~al.(2025)Li, Zhao, and Zhang]{li2025sacl}
Ming Li, Rui Zhao, and Wei Zhang.
\newblock Sacl: Understanding and combating textual bias in text-to-code retrieval, 2025.
\newblock URL \url{https://arxiv.org/pdf/2506.20081}.

\bibitem[Li et~al.(2022)Li, Choi, Gerlach, Rybkin, Chen, et~al.]{li2022competition}
Yujia Li, David Choi, Martin Gerlach, Kirill Rybkin, Xinyun Chen, et~al.
\newblock Competition-level code generation with {AlphaCode}.
\newblock \emph{Science}, 378\penalty0 (6624):\penalty0 1092--1097, 2022.

\bibitem[Liu et~al.(2024{\natexlab{a}})Liu, Xu, Deng, Yan, Li, Yin, Wang, Zeng, Wang, Wang, et~al.]{liu2024r2c2coder}
Jiaheng Liu, Zihan Xu, Zeping Deng, Bo~Yan, Qi~Li, Yining Yin, Shiqing Wang, Zhipeng Zeng, Shuo Wang, Xuying Wang, et~al.
\newblock R2c2-coder: Enhancing and benchmarking real-world repository-level code completion abilities of code llms.
\newblock \emph{arXiv preprint arXiv:2406.01359}, 2024{\natexlab{a}}.

\bibitem[Liu et~al.(2024{\natexlab{b}})Liu, Tian, Daita, Wei, Ding, Wang, Yang, and Zhang]{liu2024repoqa}
Jiawei Liu, Jia~Le Tian, Vijay Daita, Yuxiang Wei, Yifeng Ding, Yuhan~Katherine Wang, Jun Yang, and Lingming Zhang.
\newblock Repoqa: Evaluating long context code understanding.
\newblock \emph{arXiv preprint arXiv:2406.06025}, 2024{\natexlab{b}}.

\bibitem[Liu et~al.(2024{\natexlab{c}})Liu, Xu, and McAuley]{liu2023repobench_python_v11}
Tianyang Liu, Canwen Xu, and Julian McAuley.
\newblock Repobench v1.1 (python).
\newblock Dataset on Hugging Face, 2024{\natexlab{c}}.
\newblock URL \url{https://huggingface.co/datasets/tianyang/repobench_python_v1.1}.
\newblock Python portion of RepoBench v1.1, covering GitHub data from Oct 6 to Dec 31 2023; associated paper: RepoBench: Benchmarking Repository-Level Code Auto-Completion Systems (ICLR 2024).

\bibitem[Liu et~al.(2025)Liu, Zhang, Huang, and Zhao]{liu2025structural}
X.~Liu, Y.~Zhang, J.~Huang, and L.~Zhao.
\newblock Structural code search using natural language queries.
\newblock \emph{arXiv preprint arXiv:2507.02107}, 2025.

\bibitem[Liu et~al.(2024{\natexlab{d}})Liu, Lan, Hu, Liu, Zhang, Wang, Shieh, and Zhou]{liu2024codexgraphbridginglargelanguage}
Xiangyan Liu, Bo~Lan, Zhiyuan Hu, Yang Liu, Zhicheng Zhang, Fei Wang, Michael Shieh, and Wenmeng Zhou.
\newblock Codexgraph: Bridging large language models and code repositories via code graph databases, 2024{\natexlab{d}}.
\newblock URL \url{https://arxiv.org/abs/2408.03910}.

\bibitem[Liu et~al.(2024{\natexlab{e}})Liu, Yu, Liu, Chen, Cao, Ma, Zhou, and Wang]{liu2024graphcoder}
Yang Liu, Wen-Si Yu, Lemao Liu, Shuo-Yuan Chen, Shuirong Cao, Wei-Ying Ma, Huaguo Zhou, and Hao-Tian Wang.
\newblock Graphcoder: Enhancing repository-level code completion via code context graph-based retrieval and language model, 2024{\natexlab{e}}.

\bibitem[Liu et~al.(2024{\natexlab{f}})Liu, Chen, Xu, and Gao]{liu2024excs}
Zhiqiang Liu, Yongmin Chen, Jian Xu, and Jianfeng Gao.
\newblock Excs: Accelerating code search with code expansion.
\newblock \emph{Scientific Reports}, 14\penalty0 (1):\penalty0 12345, 2024{\natexlab{f}}.
\newblock \doi{10.1038/s41598-024-73907-6}.
\newblock URL \url{https://www.nature.com/articles/s41598-024-73907-6}.

\bibitem[Long et~al.(2025)Long, Zhuang, Shen, Yan, Li, and Wang]{long2025enhancing}
Xiao Long, Liansheng Zhuang, Chen Shen, Shaotian Yan, Yifei Li, and Shafei Wang.
\newblock Enhancing large language models with reward-guided tree search for knowledge graph question and answering, 2025.

\bibitem[Lozhkov et~al.(2024)Lozhkov, Li, Allal, Cassano, Lamy-Poirier, Tazi, Tang, Pykhtar, Liu, Wei, et~al.]{lozhkov2024starcoder}
Anton Lozhkov, Raymond Li, Loubna~Ben Allal, Federico Cassano, Joel Lamy-Poirier, Nouamane Tazi, Ao~Tang, Dmytro Pykhtar, Jiawei Liu, Yuxiang Wei, et~al.
\newblock Starcoder 2 and the stack v2: The next generation.
\newblock \emph{arXiv preprint arXiv:2402.19173}, 2024.

\bibitem[Ma et~al.(2024)Ma, Yang, Cao, Li, Huang, and Li]{ma2024alibaba}
Yingwei Ma, Qingping Yang, Rongyu Cao, Binhua Li, Fei Huang, and Yongbin Li.
\newblock Alibaba lingmaagent: Improving automated issue resolution via comprehensive repository exploration, 2024.

\bibitem[Mastropaolo et~al.(2021)Mastropaolo, Bavota, Linares-V{\'a}squez, Di~Penta, Oliveto, and Lanza]{mastropaolo2021empirical}
Antonio Mastropaolo, Gabriele Bavota, Mario Linares-V{\'a}squez, Massimiliano Di~Penta, Rocco Oliveto, and Michele Lanza.
\newblock An empirical study on the usage of code search in modern development.
\newblock In \emph{Proceedings of the 43rd International Conference on Software Engineering}, 2021.

\bibitem[Mou et~al.(2016)Mou, Li, Zhang, Wang, and Jin]{mou2016tbcnn}
Lili Mou, Ge~Li, Lu~Zhang, Tao Wang, and Zhi Jin.
\newblock Convolutional neural networks over tree structures for programming language processing.
\newblock \emph{arXiv preprint arXiv:1409.5718}, 2016.

\bibitem[Muennighoff et~al.(2022)Muennighoff, Tazi, Magne, and Reimers]{muennighoff2022mteb}
Niklas Muennighoff, Nouamane Tazi, Loïc Magne, and Nils Reimers.
\newblock {MTEB: Massive Text Embedding Benchmark}, 2022.

\bibitem[Nijkamp et~al.(2022)Nijkamp, Pang, Hayashi, Tu, Wang, Zhou, Savarese, and Xiong]{nijkamp2022codegen}
Erik Nijkamp, Bo~Pang, Hiroaki Hayashi, Lifu Tu, Huan Wang, Yingbo Zhou, Silvio Savarese, and Caiming Xiong.
\newblock Codegen: An open large language model for code with multi-turn program synthesis.
\newblock \emph{arXiv preprint arXiv:2203.13474}, 2022.

\bibitem[Ouyang et~al.(2024)Ouyang, Yu, Ma, Xiao, Zhang, Jia, Han, Zhang, and Yu]{repograph2025}
Siru Ouyang, Wenhao Yu, Kaixin Ma, Zilin Xiao, Zhihan Zhang, Mengzhao Jia, Jiawei Han, Hongming Zhang, and Dong Yu.
\newblock Repograph: Enhancing ai software engineering with repository-level code graph.
\newblock \emph{arXiv preprint arXiv:2410.14684}, 2024.

\bibitem[Pan et~al.(2024)Pan, Hu, Xia, and Yang]{catcoder2025}
Zhiyuan Pan, Xing Hu, Xin Xia, and Xiaohu Yang.
\newblock Enhancing repository-level code generation with integrated contextual information.
\newblock \emph{arXiv preprint arXiv:2406.03283}, 2024.

\bibitem[Parvez et~al.(2021)Parvez, Ahmad, Chakraborty, Ray, and Chang]{parvez2021retrieval}
Md~Rizwan Parvez, Wasi~Uddin Ahmad, Saikat Chakraborty, Baishakhi Ray, and Kai-Wei Chang.
\newblock Retrieval augmented code generation and summarization.
\newblock \emph{arXiv preprint arXiv:2108.11601}, 2021.

\bibitem[Robertson \& Zaragoza(2009)Robertson and Zaragoza]{robertson2009probabilistic}
Stephen Robertson and Hugo Zaragoza.
\newblock The probabilistic relevance framework: Bm25 and beyond.
\newblock \emph{Foundations and Trends® in Information Retrieval}, 3\penalty0 (4):\penalty0 333--389, 2009.

\bibitem[Roziere et~al.(2023)Roziere, Gehring, Gloeckle, Sootla, Gat, Tan, Adi, Liu, Sauvestre, Remez, et~al.]{roziere2023code}
Baptiste Roziere, Jonas Gehring, Fabian Gloeckle, Sten Sootla, Itai Gat, Xiaoqing~Ellen Tan, Yossi Adi, Jingyu Liu, Romain Sauvestre, Tal Remez, et~al.
\newblock Code llama: Open foundation models for code.
\newblock \emph{arXiv preprint arXiv:2308.12950}, 2023.

\bibitem[Silver et~al.(2017)Silver, Schrittwieser, Simonyan, Antonoglou, Huang, Guez, Hubert, Baker, Lai, Bolton, et~al.]{silver2017mastering}
David Silver, Julian Schrittwieser, Karen Simonyan, Ioannis Antonoglou, Aja Huang, Arthur Guez, Thomas Hubert, Lucas Baker, Matthew Lai, Adrian Bolton, et~al.
\newblock Mastering the game of go without human knowledge.
\newblock \emph{Nature}, 550\penalty0 (7676):\penalty0 354--359, 2017.
\newblock \doi{10.1038/nature24270}.

\bibitem[Singhal et~al.()Singhal, Ghosh, Mundra, Dadlani, and Dutta]{singhal2025code2json}
Aryan Singhal, Rajat Ghosh, Ria Mundra, Harshil Dadlani, and Debojyoti Dutta.
\newblock Code2json: Can a zero-shot llm extract code features for code rag?
\newblock In \emph{ICLR 2025 Third Workshop on Deep Learning for Code}.

\bibitem[Sun et~al.(2025)Sun, Wang, and Chen]{sun2025repo}
J.~Sun, Y.~Wang, and H.~Chen.
\newblock Repository-level code search with neural retrieval methods.
\newblock \emph{arXiv preprint arXiv:2502.07067}, 2025.

\bibitem[Sun et~al.(2020)Sun, Zhu, Xiong, Sun, Mou, and Zhang]{sun2020treegen}
Zeyu Sun, Qihao Zhu, Yingfei Xiong, Yican Sun, Lili Mou, and Lu~Zhang.
\newblock Treegen: A tree-based transformer architecture for code generation.
\newblock In \emph{Proceedings of the AAAI Conference on Artificial Intelligence}, volume~34, pp.\  8964--8971, 2020.

\bibitem[Tao et~al.(2025)Tao, Zhang, Tang, Peng, Zhu, Liu, Yang, Zhang, Xu, Zhang, Zhu, Wang, Yu, Li, and Di]{tao2025codegraphmodel}
Hongyuan Tao, Ying Zhang, Zhenhao Tang, Hongen Peng, Xukun Zhu, Bingchang Liu, Yingguang Yang, Ziyin Zhang, Zhaogui Xu, Haipeng Zhang, Linchao Zhu, Rui Wang, Hang Yu, Jianguo Li, and Peng Di.
\newblock Code graph model (cgm): A graph-integrated large language model for repository-level software engineering tasks.
\newblock \emph{arXiv preprint arXiv:2505.16901}, 2025.

\bibitem[Wang et~al.(2023{\natexlab{a}})Wang, Chen, Ren, Sun, and Zhang]{wang2023coderag}
Ke~Wang, Liang Chen, Xiang Ren, Yizhou Sun, and Yu~Zhang.
\newblock {CodeRAG}: Enhancing code large language models with retrieval-augmented generation.
\newblock \emph{arXiv preprint arXiv:2309.14509}, 2023{\natexlab{a}}.

\bibitem[Wang et~al.(2023{\natexlab{b}})Wang, Li, Qian, Yang, Wang, Shang, Kumar, Tan, Ray, Bhatia, et~al.]{wang2023recode}
Shiqi Wang, Zheng Li, Haifeng Qian, Chenghao Yang, Zijian Wang, Mingyue Shang, Varun Kumar, Samson Tan, Baishakhi Ray, Parminder Bhatia, et~al.
\newblock Recode: Robustness evaluation of code generation models.
\newblock In \emph{Proceedings of the 61st Annual Meeting of the Association for Computational Linguistics (Volume 1: Long Papers)}, pp.\  13818--13843, 2023{\natexlab{b}}.

\bibitem[Wang et~al.(2025)Wang, Wang, Niu, Lou, Qian, Wang, Tanchang, Wang, Chen, Chen, Ma, Fan, Li, Yuan, Wang, Fei, Xie, Xue, Hui, Li, Li, Liu, Qian, Li, Chen, Feng, Chen, Li, Jianhuang, Xia, Zhao, Xingzhang, and Zhou]{wang2025qwen3}
Shitong Wang, Jianing Wang, Zexuan Niu, Yichao Lou, Hongjin Qian, Xinyu Wang, Ru~Tanchang, Chengyu Wang, Cen Chen, Jinmeng Chen, Ziyang Ma, Yiming Fan, Peng Li, Zheng Yuan, Chang'an Wang, Zhaoye Fei, Ruobing Xie, Fuzhao Xue, Binyuan Hui, Yangfan Li, Jinze Li, Zhenghao Liu, Qimin Qian, Jianxin Li, Yufeng Chen, Sinan Feng, Wenhao Chen, Yanxiang Li, Jianhuang, Guisong Xia, Weilin Zhao, Xingzhang, and Jingren Zhou.
\newblock {Qwen3 Embedding: Advancing Text Embedding and Reranking Through Foundation Models}, 2025.

\bibitem[Wang et~al.(2020)Wang, Li, Ma, Xia, and Jin]{wang2020ccgraph}
Wen Wang, Guowei Li, Biao Ma, Xin Xia, and Zhi Jin.
\newblock Ccgraph: a pdg-based code clone detector with approximate graph matching.
\newblock In \emph{Proceedings of the 28th ACM Joint Meeting on European Software Engineering Conference and Symposium on the Foundations of Software Engineering}, pp.\  1115--1126, 2020.

\bibitem[Wang et~al.(2024)Wang, Liu, Chen, Liu, Lu, Wang, Liu, Liu, and Zhang]{wang2024rlcoder}
Yanli Wang, Haoyang Liu, Yang Chen, Huan Liu, Yao Lu, Yansong Wang, Xueyu Liu, Ke~Liu, and Yaqin Zhang.
\newblock Reinforcement learning for repository-level code completion.
\newblock \emph{arXiv preprint arXiv:2407.19487}, 2024.

\bibitem[Wang et~al.(2023{\natexlab{c}})Wang, Zhang, Zhang, Chen, Chen, Sun, and Zhang]{wang2023enhancing}
Yuxiang Wang, Jiaxin Zhang, Weinan Zhang, Yiming Chen, Ting Chen, Maosong Sun, and Yue Zhang.
\newblock Enhancing conversational search: Large language model-aided informative query rewriting.
\newblock In \emph{Findings of the Association for Computational Linguistics: EMNLP 2023}, pp.\  5937--5949, Singapore, December 2023{\natexlab{c}}. Association for Computational Linguistics.
\newblock URL \url{https://aclanthology.org/2023.findings-emnlp.398}.

\bibitem[Wu et~al.(2019)Wu, Petroni, Josifoski, Riedel, and Zettlemoyer]{wu2019scalable}
Ledell Wu, Fabio Petroni, Martin Josifoski, Sebastian Riedel, and Luke Zettlemoyer.
\newblock Scalable zero-shot entity linking with dense entity retrieval.
\newblock \emph{arXiv preprint arXiv:1911.03814}, 2019.

\bibitem[Ye \& Bunescu(2018)Ye and Bunescu]{ye2018queryreform}
Xiang Ye and Razvan Bunescu.
\newblock Learning to reformulate queries for code search.
\newblock In \emph{Proceedings of the 27th ACM International Conference on Information and Knowledge Management}, pp.\  1363--1372. ACM, 2018.

\bibitem[Ye et~al.(2022)Ye, Wang, Wang, He, Sun, Liu, Tang, Chen, Yin, Zhou, et~al.]{ye2022retrieval}
Ziyang Ye, Yue Wang, Shufan Wang, Bin He, Yu~Sun, Dayiheng Liu, Duyu Tang, Shujie Chen, Jian Yin, Ming Zhou, et~al.
\newblock Retrieval-augmented code generation and summarization.
\newblock In \emph{International Conference on Learning Representations}, 2022.

\bibitem[Yu et~al.(2025)Yu, Zhang, Zhao, Huang, Yao, Ding, and Zhao]{orcaloca2025}
Zhongming Yu, Hejia Zhang, Yujie Zhao, Hanxian Huang, Matrix Yao, Ke~Ding, and Jishen Zhao.
\newblock Orcaloca: An llm agent framework for software issue localization.
\newblock \emph{arXiv preprint arXiv:2502.00350}, 2025.

\bibitem[Zhang et~al.(2021)Zhang, Wang, Xu, and Lyu]{zhang2021degraphcs}
Hongyu Zhang, Xin Wang, Bowen Xu, and Michael~R Lyu.
\newblock Enhancing code search with graph neural networks.
\newblock In \emph{2021 IEEE/ACM 43rd International Conference on Software Engineering (ICSE)}, pp.\  1607--1618. IEEE, 2021.

\bibitem[Zhang et~al.(2025)Zhang, Chen, Song, and Wang]{zhang-etal-2025-rekg}
Jinhao Zhang, Zihan Chen, Xiaowei Song, and Shuhan Wang.
\newblock {REKG}-{MCTS}: Reinforcing {LLM} reasoning on knowledge graphs with monte carlo tree search.
\newblock In \emph{Findings of the Association for Computational Linguistics: ACL 2025}, Bangkok, Thailand, July 2025. Association for Computational Linguistics.
\newblock URL \url{https://aclanthology.org/2025.findings-acl.484}.

\bibitem[Zhang et~al.(2024{\natexlab{a}})Zhang, Wang, Zhang, Xu, and Sun]{zhang2024excs}
Wen Zhang, Shu Wang, Kai Zhang, Hui Xu, and Zhongyuan Sun.
\newblock Excs: Accelerating code search with code expansion.
\newblock \emph{Scientific Reports}, 14\penalty0 (1):\penalty0 23976, 2024{\natexlab{a}}.

\bibitem[Zhang et~al.(2024{\natexlab{b}})Zhang, Liu, Kong, Fu, Song, and Jiang]{zhang-etal-2024-draco}
Yifan Zhang, Zhen Liu, Li~Kong, Xipeng Fu, Linqi Song, and Jie Jiang.
\newblock {D}ra{C}o: Dataflow-guided retrieval augmentation for repository-level code completion.
\newblock In \emph{Findings of the Association for Computational Linguistics: ACL 2024}, pp.\  1469--1484, Bangkok, Thailand, August 2024{\natexlab{b}}. Association for Computational Linguistics.
\newblock URL \url{https://aclanthology.org/2024.findings-acl.126}.

\bibitem[Zhang et~al.(2022)Zhang, Wang, Wang, Ke, Ma, Xu, Zhou, Sun, et~al.]{zhang2022codexglue}
Yue Zhang, Ge~Wang, Shuo Wang, Xiaodong Ke, Shaowei Ma, Hongyu Xu, Ming Zhou, Xu~Sun, et~al.
\newblock {CodeXGLUE}: A machine learning benchmark dataset for code understanding and generation.
\newblock In \emph{NeurIPS Datasets and Benchmarks Track}, 2022.

\bibitem[Zhong et~al.(2024)Zhong, Lin, Liu, Zhou, and Chen]{zhong2024retrieval}
Zexuan Zhong, Xi~Victoria Lin, Jiaming Liu, Shuyan Zhou, and Danqi Chen.
\newblock Retrieval-augmented code generation and understanding.
\newblock In \emph{Advances in Neural Information Processing Systems}, 2024.

\end{thebibliography}
